\documentclass{vldb}

\pdfoutput=1
\usepackage{graphicx}
\usepackage{balance}  
\usepackage{url}

\usepackage{mathtools}
\usepackage{cite}
\usepackage{algorithm}
\usepackage{algorithmic}
\usepackage{subfig}

\newdef{definition}{Definition}

\usepackage[pdftex]{hyperref}
\hypersetup{%
pdfauthor={blind submission}, %
bookmarksnumbered, %
pdfstartview={c}, %
colorlinks,%
citecolor=black, %
filecolor=black, %
linkcolor=black, %
urlcolor=black}

\newcommand{\vertexlabel}{\mathit{l}}

\newcommand{\ipt}{\mathit{ipt}}
\newcommand{\str}{\mathit{str}}
\newcommand{\vmt}[1]{\mathit{VM}^{(#1)}}
\newcommand{\VMT}{\mathit{VM}}
\newcommand{\vmit}[2]{\mathit{VM}_{#1}^{(#2)}}
\newcommand{\INDSTATE}[1][1]{\STATE\hspace{#1\algorithmicindent}}
\newcommand{\TAPR}{\textit{TAPER}}
\newcommand{\Q}{${\cal Q}$}
\newcommand{\introv}{\mathit{introversion}}
\newcommand{\extrov}{\mathit{extroversion}}
\newcommand{\paths}{\mathit{paths}}

\newcounter{saveenumerate}
\makeatletter
\newcommand{\enumeratext}[1]{%
\setcounter{saveenumerate}{\value{enum\romannumeral\the\@enumdepth}}
\end{enumerate}
#1
\begin{enumerate}
\setcounter{enum\romannumeral\the\@enumdepth}{\value{saveenumerate}}%
}
\makeatother

\begin{document}


\title{TAPER: query-aware, partition-enhancement for large, heterogenous, graphs}



%
%
%
%

\numberofauthors{2}

\author{
\alignauthor Hugo Firth\\
       \affaddr{School of Computing Science}\\
       \affaddr{Newcastle University}\\
       \email{h.firth@ncl.ac.uk}
\alignauthor Paolo Missier\\
       \affaddr{School of Computing Science}\\
       \affaddr{Newcastle University}\\
       \email{paolo.missier@ncl.ac.uk}
}

\maketitle

\begin{abstract}
Graph partitioning has long been seen as a viable approach to address Graph DBMS scalability.
A partitioning, however, may introduce extra query processing latency unless it is sensitive to a specific query workload, and optimised to minimise \textit{inter}-partition traversals for that workload.
Additionally, it should also be possible to incrementally adjust the partitioning in reaction to changes in the graph topology, the query workload, or both.
Because of their complexity, current partitioning algorithms fall short of one or both of these requirements, as they are designed for offline use and as one-off operations.

The \TAPR{} system aims to address both requirements, whilst leveraging existing partitioning algorithms. \TAPR{} takes any given initial partitioning as a starting point, and iteratively adjusts it by swapping chosen vertices across partitions, heuristically reducing the probability of inter-partition traversals for a given pattern matching queries workload. Iterations are inexpensive thanks to time and space optimisations in the underlying support data structures.

We evaluate \TAPR{} on two different large test graphs and over realistic query workloads.
Our results indicate that, given a hash-based partitioning, \TAPR{} reduces the number of inter-partition traversals by $\sim 80\%$; given an unweighted Metis partitioning, by $\sim 30\%$.
These reductions are achieved within 8 iterations and with the additional advantage of being workload-aware and usable online.
\end{abstract}
\category{H.2.4}{Database Management}{Systems}
\terms{Algorithms, Performance}
\keywords{Graph databases, Graph repartitioning, Workload mining} 

\section{Introduction} \label{section:introduction}
Pattern matching queries over labelled graphs are increasingly common in many applications.
These include fraud detection \cite{Tong2007}, recommender systems \cite{Huang2002} and social analysis \cite{Chen2009} amongst others.
Such a labelled graph has the form $G=(V,E, L_V, \vertexlabel)$, where each vertex $v$ is annotated with a label $\vertexlabel(v) \in L_V$ from a predefined set $L_V$ of labels (e.g. Purchase, Person, etc\ldots{}).
In this work we address the problem of efficiently and incrementally \textbf{improving path query performance over $k-$way partitionings} of large, heterogeneous, labelled graphs.
A $k-$partitioning $P_k(G)$ of $G$ is a disjoint family of sets $\{V_1,V_2,\ldots,V_k\}$, with $V_1 \cup \dots \cup V_k = V$.
The heterogeneity of $G$ refers to the diversity in the labels $L_V$ associated to the vertices, e.g., a social graph with $L_V = \{ Person, Post\}$ is more heterogeneous than a web graph with $L_V = \{ Url \}$
%
%

%
Partitioning large graphs is a recognised approach to addressing scalability issues in graph data management.
However, if these partitionings are of a low quality then the performance of pattern matching queries (inc. path queries), greatly decreases~\cite{Mondal2012}.
Intuitively, any measure of this partitioning quality should correspond to the number of inter-partition traversals, or $\ipt$ for short, i.e., the number of times that inter-partition edges $(v_i, v_j) \in E$ with $v_i \in V_i, v_j \in V_j, i \neq j$ are traversed during query execution.
Current systems for improving graph partition quality either optimise data placement (graph partitioners) \cite{Karypis1998a, Hendrickson1995, Stanton2012, Sanders2013}, or are based on selective vertex replication \cite{Pujol2010a, Mondal2012, Yang2012}.
We will improve on the output of graph partitioners, without considering vertex replication, i.e., $V_i \wedge V_j = \emptyset$ for $i \neq j $.
Existing graph partitioners have two main drawbacks:
Firstly, due to their computational complexity~\cite{Stanton2012}, non-streaming methods~\cite{Karypis1998a, Hendrickson1995, Sanders2013} are only suitable as offline operations, typically performed ahead of analytical workloads.
For online, non-analytical workloads, they require complete re-execution, i.e., after a series of graph updates, which may be impractical~\cite{Jindal2012}.
Simpler methods, such as grouping vertices by some hash of their ids, are efficient~\cite{Mondal2012} but yield poor $\ipt$ scores when queried.
Methods meant to partition graph streams~\cite{Stanton2012, Tsourakakis2012} lie between these two extremes, both in terms of efficiency and quality.
However, they make strong assumptions about the order of graph streams and the availability of neighbourhood information for new vertices. 
As a result they are also largely confined to offline application.
%
%

\begin{figure}
\centering
\includegraphics[width=.8\columnwidth]{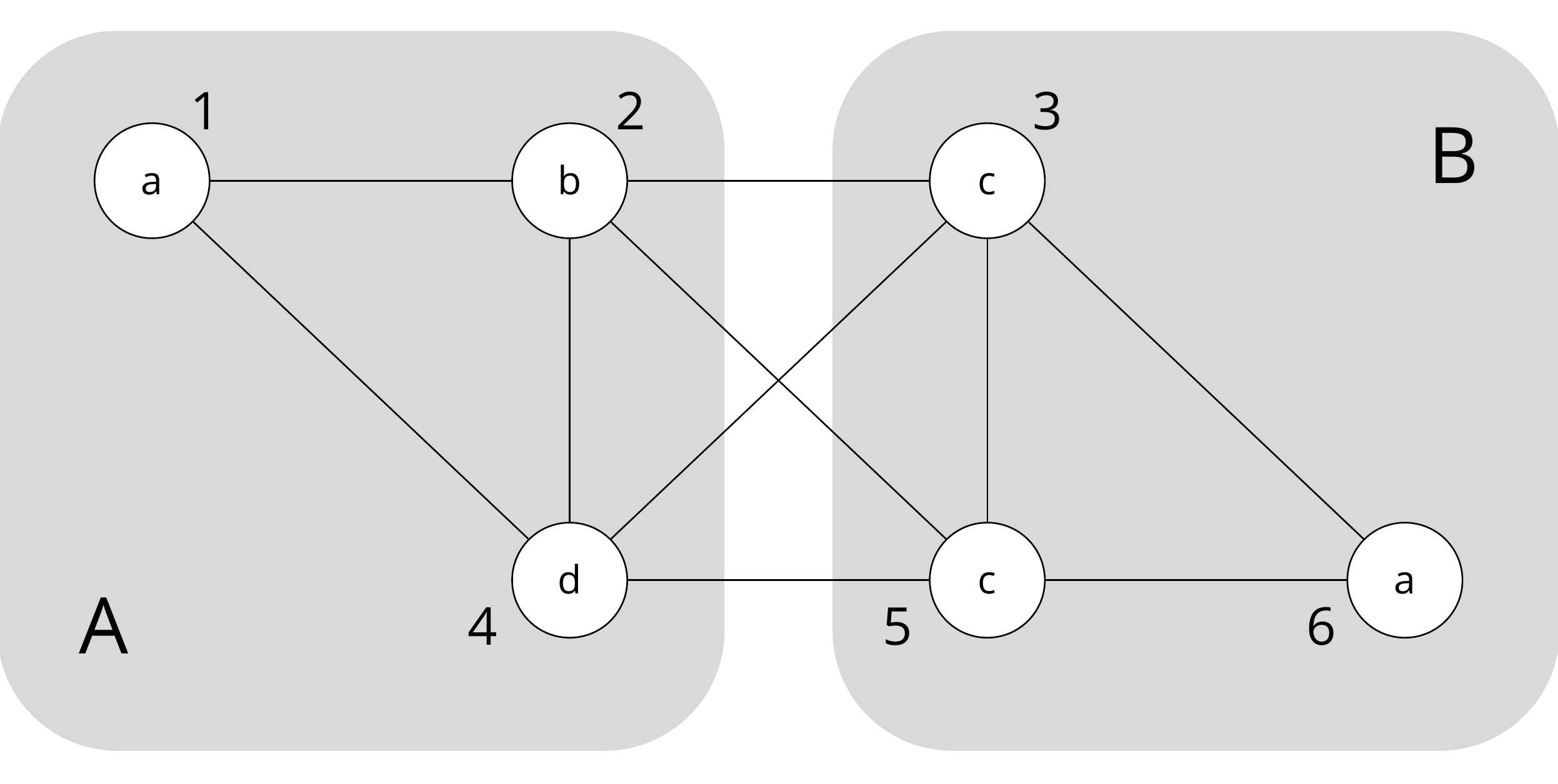}
\caption{Illustrative example graph}
\label{fig:ex-graph-1}
\end{figure}

%
Secondly, the partitioners are agnostic to query workloads as they optimise for producing the minimum number or weight of inter-partition edges (\textit{min edge-cut}).
This is equivalent to assuming uniform, or at least constant, likelihood of traversal for each edge throughout query processing.
This assumption is unrealistic as a workload may traverse a limited subset of edges, which is specific to its query patterns and subject to change.
To appreciate the importance of query-sensitive partitioning, consider the graph of Fig.\ref{fig:ex-graph-1}. 
The partitioning A and B is optimal following a balanced min edge-cut approach~\cite{Karypis1998a}, but it may not be optimal when query patterns are taken into account.

Following common practice, we express queries using a Regular Path Queries \cite{mendelzon1995finding, Barcelo2010} (RPQ) formalism, which can be expressed using a restricted form of regular path expressions over the set of vertex labels.
For example, expression $c \cdot (b | d)$ evaluates to paths $(3,2), (3,4), (5,2), (5,4)$.

Notice that computing each of these paths requires 1 $\ipt$. However, it is easy to see that with the alternative partitioning $V_1 = \{ 1,3,6\}, V_2 =\{2,4,5\}$, only paths $(3,2), (5,4)$ require traversing a partition boundary, although this partitioning is not optimal with respect to min edge-cut.
Mature research of query-sensitive database partitioning is currently confined to relational DBMS \cite{Quamar2013,Curino2010}. 

\subsection{The \TAPR{} re-partitioner}\label{subsection:taper-repart}
In this paper we present \TAPR{}, a graph re-partitioning system that is sensitive to evolving query workloads.
Let 
${\cal Q} = \{(Q_1, n_1) \dots (Q_h,n_h)\}$
denote a query workload, where $n_i$ is the relative frequency of $Q_i$ in ${\cal Q}$, and  let $P_k(G)$ be an existing partitioning of $G$. 
This could be for instance a simple hash-based partitioning, or one based on an established method such as Metis \cite{Karypis1998a} multilevel partitioning \cite{Karypis1998a} or spectral recursive octasection \cite{Hendrickson1995}.

The goal of  \TAPR{} is to \textit{enhance} $P_k(G)$, by computing a new partitioning $P_k'(G,{\cal Q})$ from $P_k(G)$ that takes \Q{} into account.
The new partitioning is obtained by swapping vertices across the partitions of $P_k(G)$, using heuristics that attempt to minimise the total \textit{probability} of $\ipt$, denoted total $\extrov$ , that occur during execution of any of the queries in \Q{}.
As this method only involves moving relatively few vertices from one partition to another, it is much less expensive than a complete re-partitioning, even after many iterations.
Furthermore, by virtue of its incremental nature, \TAPR{} is able to react to changes ${\cal Q} \rightarrow {\cal Q'}$ to the workload, by re-partitioning its own partitioning, i.e., 
\begin{equation}
P_k(G,Q) \xrightarrow{{\cal Q}'} P_k'(G,Q')
\label{eq:graph-repart}
\end{equation}
In general, given an initial, possibly workload-agnostic, and non-optimal initial partitioning $P^0_k(G)$, \TAPR{} can be used to compute a progression of partitionings:
\begin{equation}
P^0_k(G) \xrightarrow{{\cal Q}_1} P^1_k(G,{\cal Q}_{1}) \xrightarrow{{\cal Q}_2} P^2_k(G,{\cal Q}_{2}) \dots	
\label{eq:repeated-graph-repart}
\end{equation}

with the property that each $P^i_k(G,{\cal Q}_{i}) $ exhibits better quality than $P^0_k(G)$, and each is approximately optimised for the corresponding workload ${\cal Q}_i$.

\TAPR{} makes use of space-efficient main-memory data structures to encode \Q{} and to associate estimates of traversal probability with the edges in $G$.
These are then used to calculate the $\extrov$ of each vertex in its partition.
A \TAPR{} re-partitioning step, as in def.~\ref{eq:graph-repart}, is actually several internal iterations of a vertex-swapping procedure aimed at reducing $\extrov$ for each vertex in turn.

\subsection{Contributions}

Our specific contributions are as follows:
\textbf{Firstly}, from the notion of \textit{stability} of a partition \cite{Delvenne2013} we derive an operational metric of partitioning quality, expressed in terms of $\extrov$ for each vertex;
\textbf{Secondly}, we describe an encoding of traversal probabilities for each edge in $G$, given \Q{}, which is space-efficient and show how they can be updated following the evolution of \Q{};
%
\& \textbf{Thirdly}, we show how \TAPR{} makes use of these structures to iteratively achieve a re-partitioning step (def.~\ref{eq:graph-repart}).
We present an extensive evaluation of the \TAPR{} system using both real and synthetic graphs of varying sizes and compare its performance and scalability against one-off, workload-agnostic partitionings obtained using the popular Metis approach\footnote{Metis:~\url{http://bit.ly/1tqUcSQ}}, without edge weights.
In our experiments we use both a simple hash-based partitioning as well as a Metis partitioning as a starting point $P^0_k(G)$ for one invocation of \TAPR.
Our results show that such an invocation of \TAPR{} converges to a stable quality within 6-7 internal iterations, and that the resulting new partitioning $P_k(G,{\cal Q})$ exhibits 
$70\%$ quality improvement when a hash-based $P^0_k(G)$ starting point is used, and about $30\%$ improvement when using a Metis initial partitioning.
Finally, we show experimentally how the quality of a partitioning degrades following successive simulated changes in \Q{}, and how it is successfully restored by repeated invoking \TAPR{} on the current partitioning and the new workload.

\subsection{Related Work}
Two main strands of prior work are relevant to our study: $(1)$ workload aware replication and data placement in distributed databases; and $(2)$ graph partitioning.
In the context of online application, distributed database queries that traverse multiple partitions are expensive\cite{Pavlo2012}, incurring high communication cost and, in some implementations, resource contention.
In order to achieve good latencies, distributed databases must find a data placement strategy which minimises these transactions, and the overhead they cause, whilst maintaining a balanced load across all machines involved.
This is also known as reducing average query \textit{span}, i.e. the number of partitions involved in answering a query.
For graph data, \textit{balanced graph partitioning} has been exhaustively studied in literature since the 1970s \cite{Karypis1998a, Hendrickson1995, Stanton2012, Sanders2013, Tsourakakis2012}, and a number of
practical solutions are available \cite{Karypis1998a, Sanders2013}.
We do not seek a new graph partitioning algorithm; rather, to propose a workload-driven method for improving partitions that already exist.
Curino et al. \cite{Curino2010, Pavlo2012} have proposed systems to tackle the related problem of workload driven data placement in distributed RDBMS.
In particular, \textit{Schism} \cite{Curino2010} captures a query workload over a period of time, modelling it as a graph, each edge of which represents tuples involved in the same transaction.
This graph is then partitioned using existing ``one-off'' techniques to achieve a min edge-cut.
Mapped back to the original database, this partitioning represents an arrangement of records which causes a minimal number of transactions in the captured workload to be distributed.
In \textit{SWORD}, Quamar et al \cite{Quamar2013} build upon the ideas presented in \textit{Schism} \cite{Curino2010}.
They use a compressed representation of the workload graph and perform incremental re-partitioning to improve the partitioning's scalability and sensitivity to workload changes.
Although the goal of these works and our own is similar, there exist major differences in approach.
For instance, in \textit{Schism}, edges directly represent the elements accessed by a query, rather than their labels as we do.
However this fine grained approach produces very large graphs, which are expensive to both partition and store, and may impact scalability\cite{Quamar2013}.
Furthermore, these works are focused on a relational data model, where typical workloads overwhelmingly consist of short, 1-2 ``hop'' queries.
This justifies Quamar et al's simplifying decision, when repartitioning a graph, to only consider queries which span a single partition. However this assumption does not hold for general graph path and pattern matching queries.
It is unclear how \textit{SWORD}'s approach would perform given a workload containing many successions of join operations, equivalent to the traversals required for graph pattern matching.

Further prior work has focused on exploiting statistical properties of a query workload, to efficiently manage graph data through replication \cite{Pujol2010a, Mondal2012, Yang2012}.
In \cite{Yang2012}, Yang et al. propose algorithms to efficiently analyse online query workloads and to dynamically replicate ``hotspots'' (cross partition clusters of vertices which are being frequently traversed), thereby temporarily dissipating network load.
Whilst highly effective at dealing with unbalanced query workloads, Yang et al. focus solely upon the replication of vertices and edges using temporary \textit{secondary} partitions.
They do not improve upon the initial partitioning, nor do they consider workload characteristics when producing it.
This can result in replication mechanisms doing far more work than is necessary over time, adversely affecting the performance of a system.
As a result, the enhancement techniques we present here would complement many workload aware replication approaches, such as that proposed by Yang et al.

\section{Definitions}\label{section:definitions}
In a labelled graph $G = (V, E, L_V, \vertexlabel)$, function $\vertexlabel:V \rightarrow L_V$ associates a label $l(v)$ from a given set $L_V$ to each vertex $v \in V$.
A \textbf{path-query} $q$ over $G$ is a regular expression over symbols in $L_V$.
We use a type of Regular Path Queries (RPQ) \cite{mendelzon1995finding}, defined by the following expression language over $L_V$:
\begin{equation}
E ::= \tau \mid (E \cdot E) \mid (E + E) \mid (E \mid E) \mid E*
\label{eq:RPQ-grammar}
\end{equation}
where $\tau \in L_V$, and as usual ``+'' represents union, ``$\mid$'' exclusive disjunction, and ``*'' the Kleene closure operator.

Let $L(Q)$ denote the regular language defined by a query $Q$.
The result of executing $Q$ is a set of subgraphs $G_i = (V_i, E_i, L_V, \vertexlabel)$, where $V_i = \{v_{i_1} \dots v_{i_n}\} \subset  V$ consists of all and only the vertices such that $\vertexlabel(v_{i_1}) \dots  \vertexlabel(v_{i_n})$ is a valid expression in $L(Q)$.
$E_i \subset E$ is the set of edges $e\in E$ that connect the vertices $v_{i_j}$ in $G$.
Note that queries that include more complex topologies, such as branching and cycles, typically require conjunctions between expressions, or other extensions to RPQs, such as those proposed by Barcelo et al. \cite{Barcelo2010}.
These extensions are not covered by the RPQ fragment defined by expression language (\ref{eq:RPQ-grammar}), and are not within the scope of this work.
%

%
%
\subsection{Stability of a graph partitioning} \label{subsection:stability}
The broad goal of \TAPR{} is to increase the quality of a $k$-way partitioning (Sec.~\ref{subsection:taper-repart}).
Here we define the measure of partition \textit{quality} which we aim to increase.
For this, we extend the notion of partition \textit{stability}, first introduced by Delvenne et al. \cite{Delvenne2013} in the context of multi-resolution community detection in graphs;
\textit{stability} is described in terms of network flow.
The main intuition is that, when a partition is \textit{stable}, a flow that originates from a point within a partition and moves randomly along paths should be trapped within the same partition for a long time.
Time is the resolution parameter.
This concept of network flow in graphs is readily modelled as a \textit{random walk}, where discrete time $t$ is measured as the number of steps.
More precisely, the stability of a partition $S_i$ is defined as the probability that it contains the same random walker both at time $t_0$ and at time $t_0 + t$, less the probability for an independent walker to be in $S_i$ 
:
$p(S_i, t_0, t_0+t) - p(S_i, t_0, \infty )$.
Note that this definition allows for the possibility of a walker crossing multiple partition boundaries before returning to its initial partition at any time during the $[t_0, t_0 + t]$ interval.
The overall stability of a partitioning $P_k(V)$ is the sum of the stability of all partitions $S_i$ where $1\leq i\leq k$.
In other words, the greater the \textit{stability} of a partitioning, the higher the probability that a random walker, having traversed \textit{t} steps, will be in the same partition where it started.
\subsection{Workload-aware stability}\label{subsection:workload-aware-stability}
In this work, we extend stability by creating a new measure of partition quality which we will refer to as \textit{workload-aware stability}.
Our extensions are driven by two main requirements.
Firstly, \TAPR{} aims to improve the \textit{quality} of a graph partitioning by minimising the probability of expensive \textit{inter}-partition traversals, when executing a \textbf{given query workload} \Q{} (def.~\ref{eq:graph-repart} in Sec.~\ref{subsection:taper-repart}).
Using stability, which models network flow as random walkers that traverse paths in a graph, gives us more flexibility than other measures of partition quality, such as \textit{edge-cut}, when we try to incorporate information on a query workload.
Stability's `walkers', represented by the probabilities in a transition matrix, may be modified to account for the specific graph patterns associated with the queries in \Q{}, along with their relative frequency.
This will reveal different dominant traversal patterns and produce a measure of quality more closely correlated with the cost of executing \Q{} over a particular graph partitioning.
Secondly, the current definition of stability as given above is also limited, as it does not account for the probability that a walker crosses partition boundaries multiple times within $t$ steps.
In contrast, we need to be able to estimate the probability that the walker \textbf{does not leave the partition} within the interval.
\subsection{The Visitor Matrix: Non-random walks with memory}\label{sec:visitor-matrix}
\begin{figure}
	\centering
	\includegraphics[width=0.95\columnwidth,height=0.8\columnwidth,keepaspectratio]{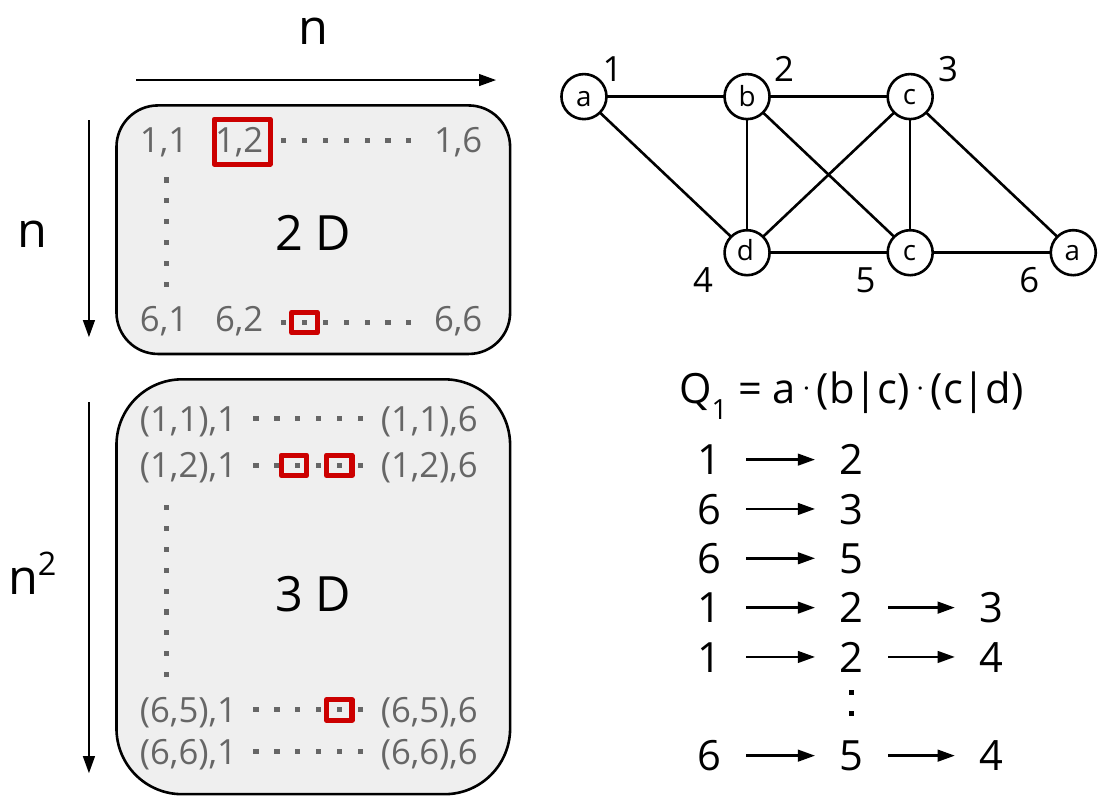}
	\caption{Visitor Matrix structure}
	\label{fig:vm-structure}
\end{figure}

We address both requirements by extending the well-known notion of a \textit{biased} random walk over a graph.
Rather than uniform transition probabilities, such a ``random'' walk assumes the more general Markov property; that is, the probability of a transition from vertex $v_k$ to $v_j$ only depends on the prior probability of being in $v_k$:
\[ Pr(v_k \rightarrow v_j| v_i \rightarrow \ldots \rightarrow  v_k) = Pr(v_k \rightarrow v_j| v_k) \]
In this case, the probabilities $Pr(v_k \rightarrow v_j| v_k)$ are captured by a transition matrix $M$:
\[ M[k,j] = Pr(v_k \rightarrow v_j | v_k) \]
and the probability of a $t$-steps walk from $v_k$ to $v_j$ is computed as $M^t[k,j]$.
However, taking into account the query matching patterns as per our requirements above, invalidates the Markov property, because the probability of a transition $v_k \rightarrow v_j$ now depends on the specific path through which we arrive at $v_k$:
\[ Pr(v_k \rightarrow v_j| p \rightarrow  v_k) \neq Pr(v_k \rightarrow v_j| p' \rightarrow v_k) \]
in general, for any two paths $p \neq p'$ leading to $v_k$.
In other words, in order to account for query matching patterns of length up to $t$, where $t$ is defined by the query expressions in ${\cal Q}$, we use a \textit{multi-step} (non-random) walk model over the graph, which has  memory of the last $t$ steps.
Each transition probability  $v_k \rightarrow v_j$ is now explicitly conditioned on the paths, of length up to $t$, which lead to $v_k$.
To represent these probabilities, we extend $M$ to a set:
\begin{equation}
\VMT(t) \equiv \{\vmt{1}, \dots, \vmt{t}\}
\label{eq:VM}
\end{equation}
of matrices, where the parameter $t$ denotes the longest query matching pattern in ${\cal Q}$, and $\vmt{k}$ has dimension $1 \leq k \leq t$.
We use the term  \textbf{Visitor Matrix} to refer to (\ref{eq:VM}).
The definition is by induction, where the base cases are the prior probabilities $Pr(v_{i})$ to be in $v_i$, for $\vmt{1}$, and the normal transition matrix $M$, for $\vmt{2}$.
Formally:
\begin{flalign*}
\vmt{1}[i]     &= Pr(v_{i})  \\
\vmt{2}[i_1, i_2] & = Pr(v_{i_1} \rightarrow v_{i_2} | v_{i_1}) = M[i_1, i_2]
\end{flalign*}
%
%
\begin{flalign}
\vmt{k}[i_1,.., i_k] = Pr(v_{i_{k-1}} \rightarrow  v_{i_{k}} | v_{i_{1}} \rightarrow \ldots \rightarrow v_{i_{k-1}})&&
\label{equation:vm-cell-to-prob}
\end{flalign}
for $2 < k \leq t$.
Fig.~\ref{fig:vm-structure} shows a representation of a Visitor Matrix with $t=3$, using a 2-dimensional matrix layout where $\vmt{3}$ is ``appended'' to $\vmt{2}$.
The cells in the matrix store probabilities for paths in the example graph to the right (originally Fig.~\ref{fig:ex-graph-1}), relative to query expression $Q_1$.
For example, path $1 \rightarrow 2 \rightarrow 3$ is an instance of query pattern $abc$, and its probability is stored in $\vmt{3}[1,2,3]$ (similarly for the other highlighted elements in the matrix).
A $\VMT$, like any finite transition matrix, is right-stochastic, i.e., each row sums to 1, and the cells represent all paths up to length $t$.
We show how compute the elements of $\VMT(t)$ for a given query workload \Q{} in section \ref{subsection:computing-vm-cells}.
In practice, $\VMT(t)$ can be partitioned into $n$ sub-matrics $\VMT_i(t)$, one for each of $n$ partitions, because we can find a permutation of the rows and columns of $\VMT$ such that $\VMT_i(t)$ is a contiguous sub-matrix of $\VMT$.
Thus, in the following we use $\VMT_i(t)$ to refer the $VM$ for partition $V_i$.
Note that the visitor matrix is impractically large to compute, with a space complexity of  $O(|V|^t)$.
In Sec. \ref{subsection:space-complexity-heuristics} we present heuristics that are designed to reduce both space complexity, as well as to avoid computing some of the cells in the $\VMT$.

\section{Enhancing a Partitioning}\label{section:model}
We are going to exploit the VM structure just defined, to compute a new partitioning $P(G, {\cal Q})$ from a partitioning $P(G)$, as in eqn.~\ref{eq:graph-repart}.
First we identify a set of vertices in each partition with highest likelihood of being the source of \textit{inter}-partition traversals ($\extrov$). 
Subsequently we swap such high-\textit{extroversion} vertices between partitions, internalising the common traversal paths resulting from ${\cal Q}$ in single partitions. 
As we will show experimentally, repeated iterations of these steps reduce the overall likelihood of inter-partition traversal across all partitions $V_i$, and thus, indirectly, increase \textit{workload-aware stability}\footnote{We never explicitly calculate stability, as it is an expensive global measure 
, unsuitable for use as a cost function.}.
These iterations constitute one invocation of the \textit{TAPER} algorithm; not to be confused with repeated invocations given a changing workload (eqn.~\ref{eq:repeated-graph-repart}).
\subsection{Increasing stability by Vertex swapping}\label{subsection:partition-enhancement}
Informally, we define the \textit{extroversion} of a vertex $v$ to be the likelihood that it is the source of an \textit{inter}-partition traversal, given any of the query patterns in ${\cal Q}$.
\TAPR{} seeks to enhance a partitioning by determining a series of vertex swaps between graph partitions such that their total \textit{extroversion} is minimised.
This is an extension of the general graph partitioning problem, a classic approach to which is the algorithm KL/FM, proposed by Kernighan and Lin \cite{Kernighan1970} and later improved upon by Fiduccia and Mattheyes \cite{Fiduccia1982}.
They present techniques that attempt to find sets of vertices and edges which, when moved between two halves of a graph bisection, produce an arrangement that is globally optimal for some criteria (usually min edge-cut).
Karypis and Kumar \cite{Karypis1998a} subsequently generalise this technique to address the problem of \textit{k}-way partitioning, in an algorithm which they call \textit{Greedy Refinement}.
Greedy Refinement selects a random boundary vertex\footnote{A vertex with neighbours in $\geq 1$ external partitions.} and orders the partitions to which it is adjacent by the potential \textit{gain} (reduction in edge-cut) of moving the vertex there, subject to some partition balance constraints.
If a move does not satisfy chosen balance constraints, progressively less beneficial destination partitions are considered.
Finally, the move will be performed.
Greedy Refinement has been shown to converge within 4-8 iterations.
It is this algorithm which we use as the basis for our \TAPR{}'s vertex swapping procedure.
However, rather than reduction in edge-cut, we use the reduction in \textit{extroversion} as our measure of gain for evaluating vertex swaps.
There are some other key differences between our own approach, and that of Greedy Refinement.
Firstly, Greedy Refinement considers vertices at random from the boundary set, whilst we consider only the set of most \textit{extroverted} vertices, in descending order of \textit{extroversion}.
This reduces the number of swaps performed and so should improve performance.
Secondly, Greedy Refinement is designed to operate on a graph compressed using a \textit{matching} algorithm, so every vertex move corresponds to the movement of a cluster of vertices in the original graph.
Without this trait, Greedy Refinement would be more susceptible to being trapped in local optimisation minima: as vertex clusters are iteratively moved across partition boundaries edge-cut may temporarily increase.
We do not operate on a compressed graph; instead we opt for a simple flood fill approach, detailed in section \ref{subsection:vertex-swapping}.
Using traversal probabilities, precomputed in the visitor matrix, we identify a vertex $v$'s \textit{family}: those vertices likely to be the source of traversals to $v$.
This is the clique of vertices which should accompany a swapping candidate to a new partition.
\subsection{Introversion and Extroversion}\label{subsection:introversion-extroversion}
We now formally define a vertex's \textit{introversion} (and, symmetrically, its \textit{extroversion}), in terms of the VM.
Given $v \in V_i$, we have seen that a VM cell
$\vmit{i}{k+2}[v_{i_1}, \ldots, v_{i_{k}}, v, v']$
denotes the probability of a transition from $v$ to $v'$, given a path $p = v_{i_1} \rightarrow \ldots \rightarrow v_{i_k} \rightarrow v$ that matches a query pattern.
Let $\paths(v, V_i)$ denote the set of all such paths in $V_i$, i.e. those that match a query pattern in ${\cal Q}$ and end in $v$.
We defined the $\introv(v)$ of $v$ as the total probability of such transition occurring, summed over every path $p \in \paths(v, V_i)$ and every destination vertex $v' \in V_i$.
Formally:
\begin{align}
\introv(v) &= \frac{1}{Pr(v)} \sum_{p} (\ Pr(p)\cdot \sum_{v' \in V_i} \VMT_{i}(t)[p,v']) \nonumber \\
           &\text{for all } p \in \paths(v, V_i)
\label{eq:introv}
\end{align}
where for path $p= v_{i_1} \rightarrow \ldots \rightarrow v_{i_k} \rightarrow v$ of length $k+1$, we have:
\begin{align*}
	Pr(p)& = Pr(v | v_{i_1} \rightarrow \ldots \rightarrow v_{i_k})\cdot \\
 	      & Pr(v_{i_k} | v_{1} \rightarrow \ldots \rightarrow v_{i_{k-1}})\cdot \ldots \cdot Pr(v_{i_2} | v_{i_1}) \cdot Pr(v_{i_1}) = \\
 	      &\vmit{i}{k+1}[v_{i_1},\ldots ,v] \cdot \\
          &\vmit{i}{k}[v_{i_1},\ldots, v_{i_k}] \cdot \ldots \cdot \vmit{i}{2}[v_{i_1},v_{i_2}] \cdot \vmit{i}{1}[v_{i_1}]
\end{align*}
and the total \textit{intra}-partition traversal probability is divided by the total probability of all traversal paths to $v$:
\[ Pr(v) = \sum_{p \in \paths(v, V_i)} Pr(p)\]
to account for the percentage of the traversals from $v$ that are internal.
Symmetrically, we define the \textit{extroversion} of vertex $v$ as the total likelihood of inter-partition traversal $v \rightarrow v'$, where $v \in V_i$ and $v' \in V_j$, $j\neq i$.
As the VM is stochastic and we may assume that a partition's VM forms a sub-matrix of the global VM, \textit{inter}-partition probabilities are the complement to 1 of the \textit{intra}-partition probabilities:
\begin{align}
\extrov(v) &= \frac{1}{Pr(v)} \sum_{p} (\ Pr(p)\cdot (1 - \sum_{v' \in V_i} \VMT_{i}(t)[p,v'])) \nonumber \\
		   &\text{for all } p \in \paths(v, V_i)
\label{eq:extrov}
\end{align}

\section{Prefix Trie encoding of query expressions}\label{sec:trie-encoding}

We use a prefix trie, which we have called the Traversal Pattern Summary Trie (\textit{TPSTry}), to encode the set of path expressions defined by each new query $Q$ in our workload ${\cal Q}$.
Combined with continuous tracking of query frequencies over a time window $t$, the TPSTry gives us a compact way to represent legal paths that may lead to each vertex $v$ in $G$, along with each path's current probability of being traversed.
From the stream of regular expressions which comprise the query workload ${\cal Q}$, we derive a dictionary set $D$ of all label sequences (strings) described by these expressions.
If a sequence of vertices $p$ is \textit{connected}, such that $(p_n, p_{n-1}) \in E$, and its corresponding sequence of labels $l(p)$ is a prefix of some sequence from $D$, then that sequence is considered \textit{legal}.
A trie is highly efficient at matching prefixes for multiple sequences or strings.
The idea of using a trie is inspired by Li et al. \cite{Li2009} who use them to encode sequences of clicked hyperlinks over a web graph, summarising the top $k$ most frequent patterns in web browsing sessions.
In our context, a sequence of clicked hyperlinks is just a particular case of generic traversals over more general forms of graph data.
%
%
%

\begin{figure}
    \parbox{0.40\columnwidth}{
    	\centering
    	\includegraphics[width=.32\columnwidth]{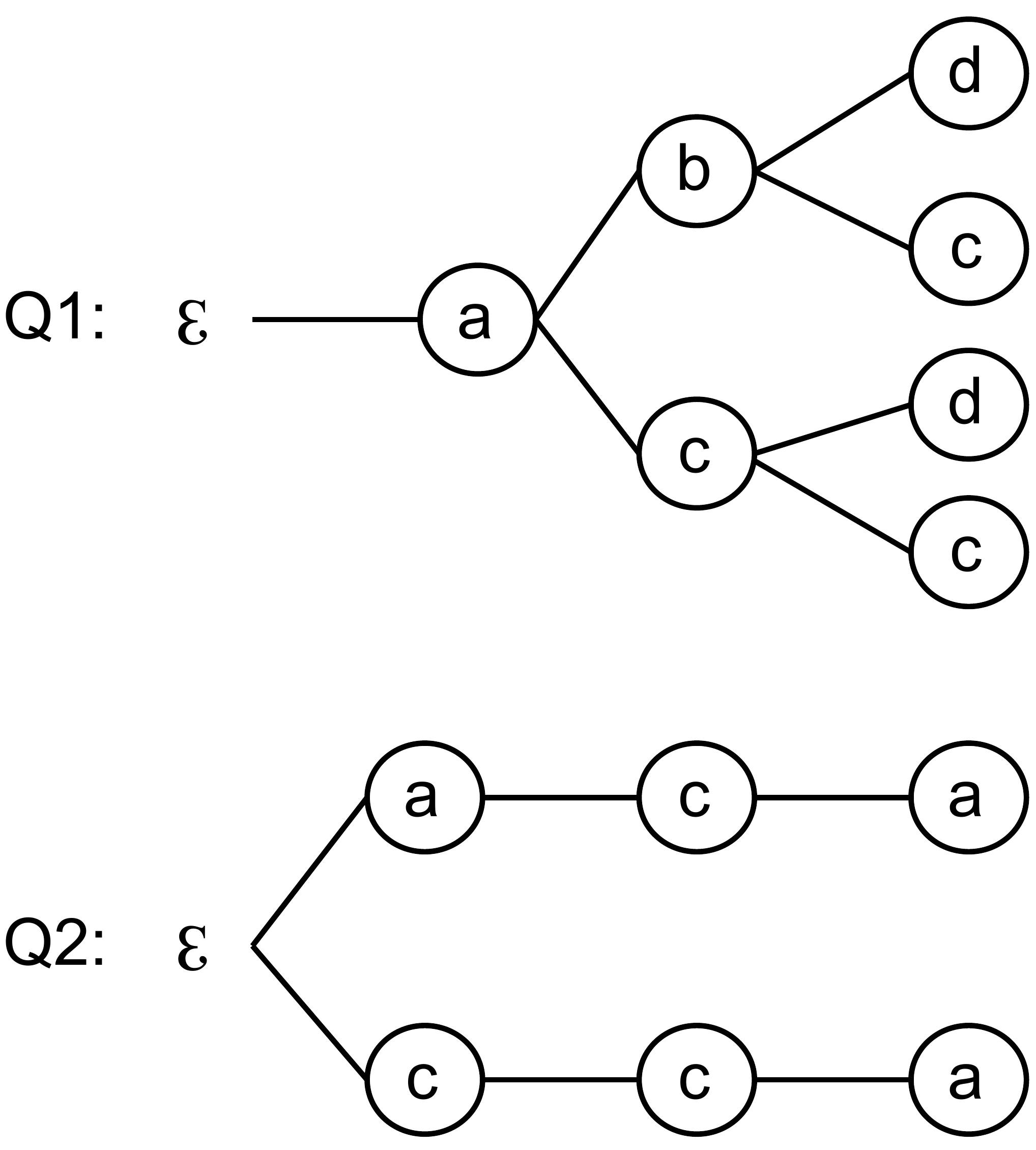}}
    \quad
    \quad
    \parbox{0.50\columnwidth}{
        \centering
        \includegraphics[width=.44\columnwidth]{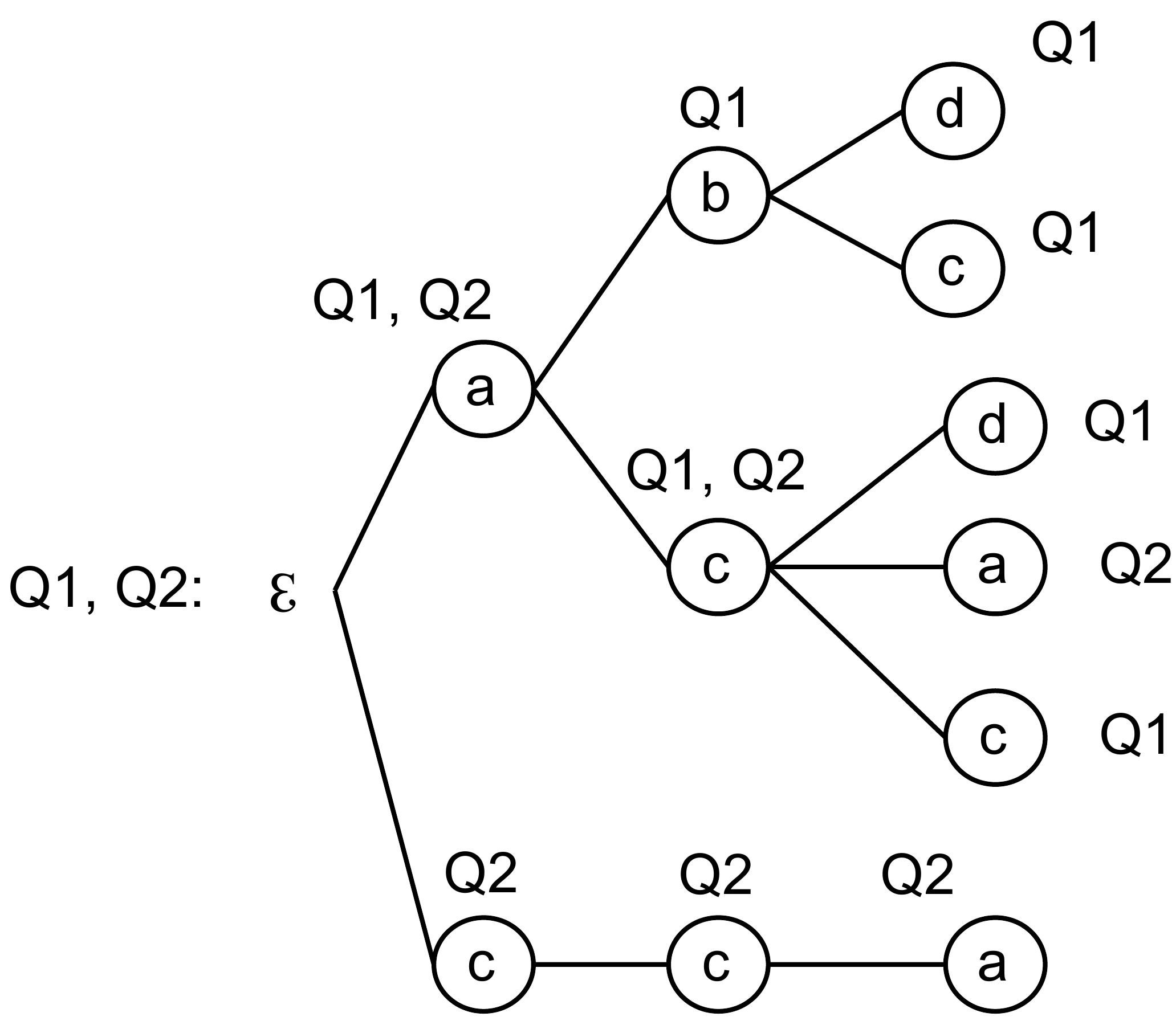}}
            \caption{Summary trie construction from queries.}
            \label{fig:ex-trie-1}
\end{figure}

%
Instead of encoding all actual graph traversals, however, we only encode query patterns in terms of the labels associated with each vertex.
Then we associate probabilities to each node in this, smaller, trie of labels.
In practice, each path in the trie is an intensional representation of a (possibly very large) set of paths in the graph, namely those whose vertices match the sequence of labels in the trie branch.
This representation is very compact, because this trie grows with $|L_V|^t$, where $t$ is the length of the longest path expressed by queries in ${\cal Q}$ and $L_V$ is typically small.
Of course, one path in the trie now corresponds to a set of paths in the graph.
We are going to take this one-many relationship into account when we convert the probabilities associated with nodes in the trie, into the probabilities associated with vertices in the graph, i.e., the elements of the VM.
Given a workload ${\cal Q}$, \textit{TPSTry} is constructed by mapping each new regular expression $Q \in {\cal Q}$ to a set of strings, and adding these to a trie using standard trie insertion procedure.
Each node in the \textit{TPSTry} which corresponds to one of these added strings is then labelled with the expression $Q$, even if the node existed as the result of a distinct previous expression\footnote{\textit{TPSTry} nodes may be labelled with multiple queries}.
The labels for each query in ${\cal Q}$ are hashes of the expressions themselves, as these are guaranteed to be unique\footnote{We use $Q_i$ labels in examples for readability}.
If an expression is not seen within the preceding time $t$ (i.e. has a frequency of 0) then it label is removed from nodes in the trie; any node without \textbf{any} query labels is also removed.
Such an infrequent expression is then treated as new in future.
The mapping $s = \str(Q)$ of a query expression $Q$ to string $s$ is straightforward and is defined as follows ($\mathit{append}(x,y)$ simply appends string $y$ to string $x$):
\begin{align*}
&\str(l) = \{ l \} \text{ for each } l \in L_V\\
&\str(e_1 \mid e_2) = \str(e_1) \cup \str(e_2) \\
&\str(e_1 \cdot e_2) = \{\mathit{append}(x,y) | x \in \str(e_1), y \in \str(e_2) \}\\
&\str(e^N) = \str(e \cdot e \ldots e)  \quad \text{// $N$ times}
\end{align*}
\textbf{Example}.
Consider again the graph in Fig. ~\ref{fig:ex-graph-1} and the expressions $Q_1 = a\cdot (b|c)\cdot (c|d)$ and $Q_2 = (c|a)\cdot c\cdot a$.
These two expressions are encoded using the two prefix trees  in Fig.~\ref{fig:ex-trie-1} (a). The two trees are then further combined into the single prefix tree in Fig.~\ref{fig:ex-trie-1}(b), with each node labelled with the set of queries it pertains to.
\subsection{Associating probabilities to trie nodes}\label{subsection:trie-probabilities}
Given a trie, such as in Fig. \ref{fig:ex-trie-1}(b), we associate a probability to each node in the trie, reflecting the relative likelihood that a sequence of vertices with those labels will be traversed in the graph.
These probabilities are periodically (re)calculated by considering both the individual contribution of each query $Q$ to the trie structure, as well as the frequency with which $Q$ appears in the workload during some preceding time $t$.
To understand these calculations, consider again Fig.~\ref{fig:ex-trie-1}, where we assume that $Q_1$, $Q_2$ each occur once in ${\cal Q}$ over time $t$, i.e., they have the same relative frequency.
Starting from root $\mathcal{E}$, consider transition $\mathcal{E} \to  a$. Its probability can be expressed as:
\[ Pr(\mathcal{E} \to  a) = Pr(\mathcal{E} \to  a |Q_1) \cdot Pr(Q_1) + Pr(\mathcal{E} \to  a |Q_2) \cdot Pr(Q_2)\]
where the conditional probabilities are computed using the labels on the nodes and the $Pr(Q_i)$ are the relative frequencies of the $Q_i$.
In the example we have $Pr(Q_1) = Pr(Q_2) = .5$, $Pr(\mathcal{E} \to  a |Q_1) = 1$ because $a$ is the only possible first match in $Q_1$'s pattern, and $Pr(\mathcal{E} \to  a |Q_2) = .5$ because initially $Q_2$ can match both $a$ and $c$, with equal probability.
Thus, $Pr(\mathcal{E} \to  a) = 1 \cdot .5 + .5 \cdot .5 = .75$.
We can now use $Pr(\mathcal{E} \to  a)$ to compute $Pr(\mathcal{E} \to a \to b)$ and $Pr(\mathcal{E} \to a \to c)$:
\begin{align*}
Pr(\mathcal{E} &\to a \to b) = \\
& Pr(\mathcal{E} \to a \to b | Q_1)\cdot Pr(Q_1) +\\
& Pr(\mathcal{E} \to a \to b | Q_2) \cdot Pr(Q_2)
\end{align*}
where
\begin{align*}
Pr(\mathcal{E} &\to a \to b | Q_1) = \\
& Pr(a \to b | \mathcal{E} \to a, Q_1) \cdot Pr(\mathcal{E} \to a | Q_1)  = \\
& .5 \cdot 1 = .5
\end{align*}
and $Pr(\mathcal{E} \to a \to b | Q_2) = 0$ because pattern $\mathcal{E} \to a \to b$ is not feasible for $Q_2$.
Thus,  $Pr(\mathcal{E} \to a \to b) = .5 \cdot .5 = .25$.
Formally, we identify each node $n$ in the trie by the sequence of $k$ steps $(n_1,n_2,\ldots,n_k)$ required to reach it from the root node, $\mathcal{E}$.
A \textit{probability label} $p(n)$ is then associated with each node, its value computed as follows:
\begin{align*}
	p(n) &=\ Pr(\mathcal{E} \to \ldots \to n_k) = \\
	& \sum_{Q_i \in {\cal Q}} Pr(\mathcal{E} \to \ldots \to n_k | Q_i) \cdot Pr(Q_i)
\end{align*}

The individual terms of the sum are conditional probabilities over the path in the trie to node $N$.
As we have seen in the example, these conditional probabilities over the paths are computed recursively on the length $k$:
\begin{align*}
Pr(\mathcal{E} &\to \ldots \to n_{k-1} \to n_k | Q_i) = \\
&  Pr(n_{k-1} \to n_k | \mathcal{E} \to \ldots \to n_{k-1}, Q_i) \cdot \\
&  P(\mathcal{E} \to \ldots \to n_{k-1}| Q_i)
\end{align*}
\begin{figure}
    \centering
    \begin{minipage}{.51\columnwidth}
    		\tiny
            $\bordermatrix{
                           & v_1         & v_2         & v_3    & v_4    & v_5    & v_6         \cr
              \ \ v_1      & 0.67        & 0.33        & 0      & 0      & 0      & 0           \cr
              \ \ v_2      & 0           & 0           & 0      & 0      & 0      & 0           \cr
              \ \ \ \vdots & \vdots      & \vdots      & \vdots & \vdots & \vdots & \vdots      \cr
              \ \ v_6      & 0           & 0           & 0.33   & 0      & 0.33   & 0.33        \cr
              v_1,v_1      & 0           & 0           & 0      & 0      & 0      & 0           \cr
              v_1,v_2      & 0           & 0           & 0.25   & 0.5    & 0.25   & 0           \cr
              \ \ \ \vdots & \vdots      & \vdots      & \vdots & \vdots & \vdots & \vdots      \cr
              v_6,v_3      & 0           & 0           & 0      & 0.25   & 0.25   & 0.5         \cr
              \ \ \ \vdots & \vdots      & \vdots      & \vdots & \vdots & \vdots & \vdots      \cr
              v_6,v_6      & 0           & 0           & 0      & 0      & 0      & 0           \cr
            }$
    \end{minipage}
    \qquad
    \parbox{0.4\columnwidth}{
        \centering
        \includegraphics[width=.4\columnwidth]{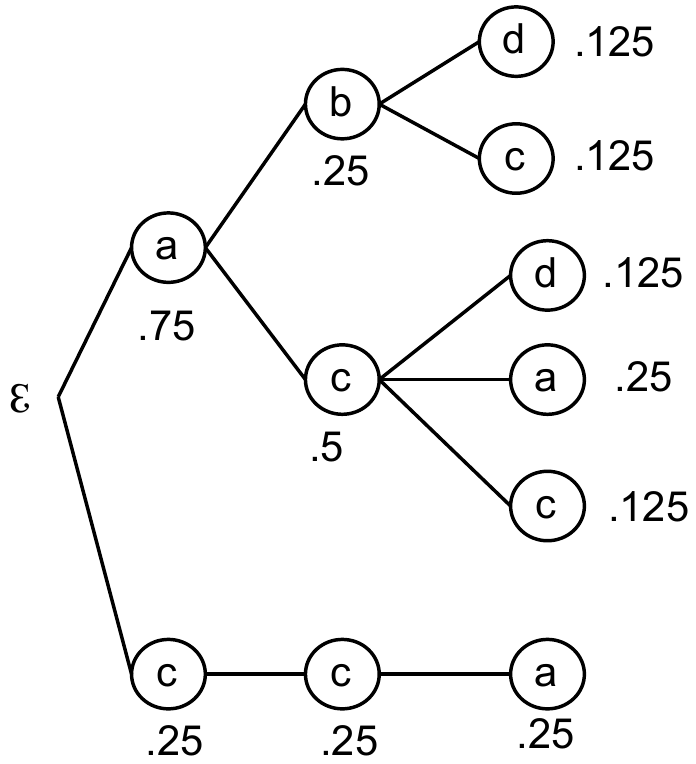}}
            \caption{Visitor Matrix (left), TPSTry probabilities (right)}
            \label{fig:visitor-matrix}
\end{figure}
\subsection{Computing VM cells with the TPSTry}\label{subsection:computing-vm-cells}
The \textit{TPSTry} encodes the current likelihood of traversing from a vertex with some label, to any connected vertex with some other label (Sec.~\ref{subsection:trie-probabilities}).
This is an abstraction over the values we actually need for the visitor matrix, which are vertex-to-vertex transition probabilities.
We may derive the desired vertex transition probabilities, given a path of previously traversed vertices $p = {p_1, p_2, \ldots, p_k}$.
First we look up the the path's corresponding sequence of vertex labels in the pattern summary trie.
This returns a set of child trie nodes $n \in N$ which represent legal labels for the next vertex to be traversed, along with each label's associated probability $p(n)$.
Subsequently, the traversal probabilities for each label are uniformly distributed amongst those neighbours of $p_k$ which share that label.
This produces a vector of traversal probabilities, one for each neighbour of the $p_k$.
This vector corresponds to a row in the visitor matrix.
For each path of traversals with length < t, the VM assumes that a subsequent traversal is guaranteed, i.e. the total traversal probability in each row is 1, and the VM is stochastic.
In reality some paths of traversals must have a total length < t, either because a query expression defines a path of a shorter length, or because a vertex does not have a neighbour with the label  required by a query expression.
A query execution engine would stop traversing in such a scenario.
We represent this non-zero probability of no subsequent traversal from a vertex as probability to traverse to the same vertex\footnote{We do not consider the possibility of self-referential edges; any probability to remain in the same vertex is equivalent to probability of no subsequent traversal.}, as this is equivalent to intra-partition traversal probability.
In Sec.\ref{sec:visitor-matrix} we described a VM cell as containing the probability of traversing to a vertex $v$ given some preceding sequence of traversals $p_1 \rightarrow p_2 \rightarrow \ldots \rightarrow p_{t-1}$.
Formally, we compute the value of a cell $\vmt{t}[p_1, \ldots, p_{t-1}, v]$ as
\begin{align*}
Pr(p_{t-1} &\rightarrow v | p_1 \rightarrow \ldots \rightarrow p_{t-1}) = \\
& Pr(\vertexlabel(p_{t-1}) \rightarrow \vertexlabel(v) | \mathcal{E} \rightarrow \vertexlabel(p1) \rightarrow \ldots \rightarrow \vertexlabel(p_{t-1}))\ \cdot \\
& Pr(p_t = v | \vertexlabel(p_t) = \vertexlabel(v), p_t\in N_G(p_{t-1}))
\end{align*}
where $\vertexlabel: V \rightarrow L_V$ is the labelling function for a graph $G$, and $N_G: V \rightarrow V$ corresponds to the set of neighbours of $v$ such that $(v, n) = e\in E$ for all $n\in N_G(v)$.
The latter term of this definition uniformly distributes the traversal probability to a vertex with label $l$ across all of of $p_{t-1}$'s $l$ labelled neighbours.
\textbf{Example}. Given the graph in Fig.~\ref{fig:ex-graph-1}, consider the element $\vmt{3}[1,2,j]$ in its visitor matrix.
The probability to be in vertex 2, having previously been in vertex 1, is given by the matrix's $\vmt{2}[1,2]^{th}$ element.
The labels of vertices 1 and 2 are $a$ and $b$ respectively.
There exist two valid suffixes to the label sequence $a\rightarrow b$: $c$ and $d$.
From the query pattern summary trie in Fig.~\ref{fig:visitor-matrix}, we know that the relative frequency of $c$ from $a\rightarrow b$ is 0.5 .
\begin{displaymath}
	Pr(b\rightarrow c | a\rightarrow b) = \frac{0.125}{0.25} = 0.5
\end{displaymath}
The relative frequency of $d$ from $a\rightarrow b$ is also 0.5 .
Vertex 2 has the neighbours 1,3,4 and 5 with the labels $a,c,d$ and $c$ respectively.
As an example, the probability of traversing to vertex 3 is the probability of traversing to a $c$ labelled vertex, divided by the number of $c$ labelled neighbours of 2.
\begin{align*}
	\vmt{3}[1,2,3] &= 0.5 \cdot Pr(j = 3 | l(j) = c, j\in N_G(2)) \\
	&= 0.5\cdot 0.5 = 0.25
\end{align*}
Therefore, as shown in Fig.~\ref{fig:visitor-matrix}(left), we have $\vmt{3}[1,2,*] = (0,0,0.25,0.5,0.25,0)$.
In the previous section (Sec.~\ref{subsection:trie-probabilities}) we mention that \textit{TPSTry} probabilities are periodically updated to reflect query frequencies changing over time.
We do not recompute VM cells for each change to the \textit{TPSTry}, instead they are lazily re-evaluated each time a vertex swapping iteration(Sec.~\ref{subsection:partition-enhancement}) is triggered.
Additionally, we store a snapshot of the \textit{TPSTry} at the point of the pervious vertex swapping iteration; if a trie node's probability remains the same between two iterations, we are able to safely avoid recomputing its associated VM cells.

\section{Implementation}\label{section:implementation}
The \TAPR{} system consists of a main algorithm for calculating elements of the Visitor Matrix and for deriving the most \textit{extroverted} vertices for each vertex swapping iteration.
The system also implements the \textit{TPSTry} traversal pattern summery trie.
In this section we present the \TAPR{} prototype architecture, we discuss heuristics for managing the space and time complexity associated with the Visitor Matrix, and we describe in detail the vertex ranking and swapping algorithm that takes place at each iteration.
\subsection{Architecture} \label{sec:architecture}
\begin{figure}
\centering
\includegraphics[scale=0.35]{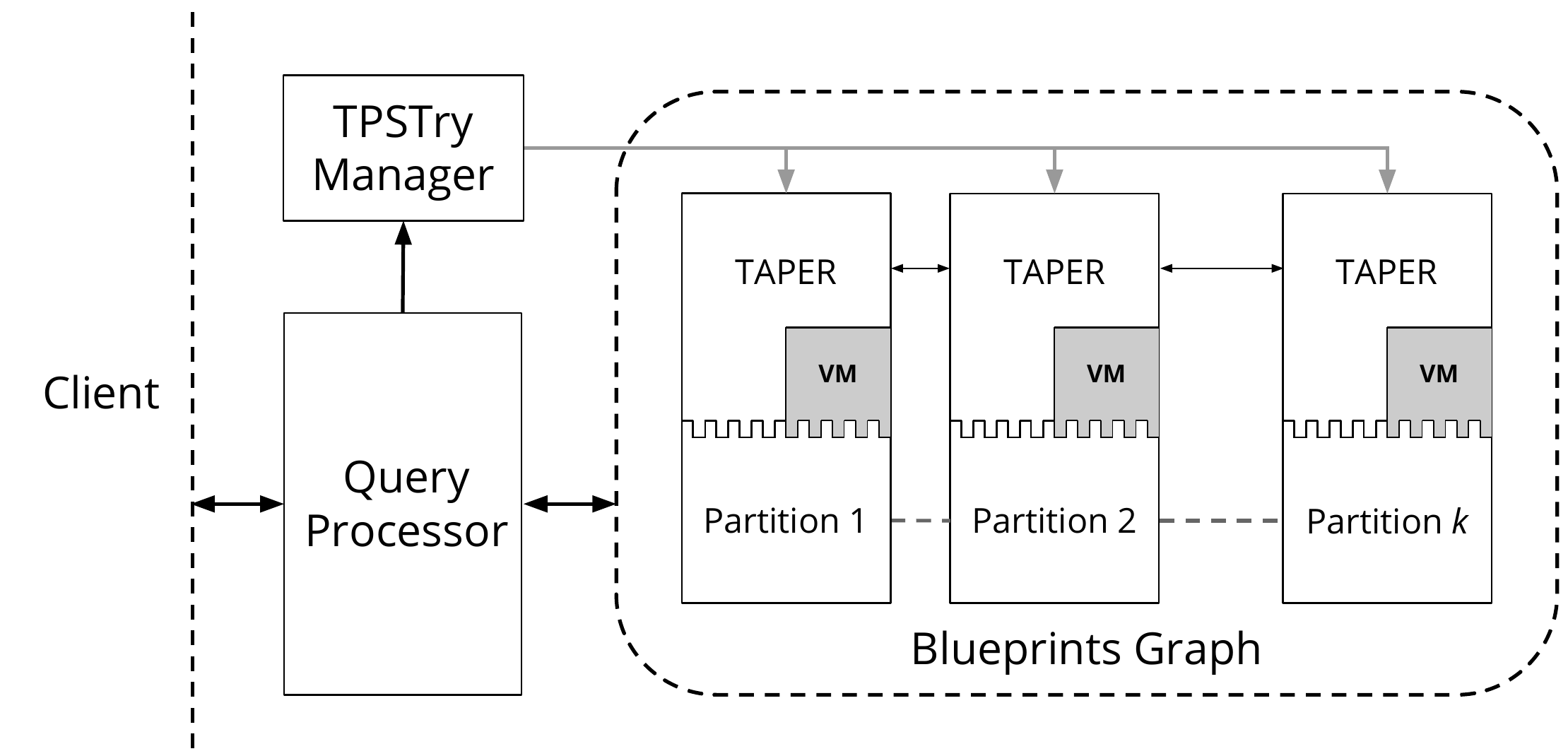}
\caption{Architecture}
\label{fig:architecture}
\end{figure}

A \TAPR{} invocation takes a partitioned graph $G$, along with a query workload ${\cal Q}$, as input and produces a new partitioning of $G$ with better \textit{workload-aware stability} (Sec.~\ref{subsection:workload-aware-stability}).
We have implemented a system prototype on top of the Tinkerpop graph processing framework\footnote{The Tinkerpop project: \url{http://bit.ly/1WNJ7HW}}, which allows us to use any of several popular GDBMS to store $G$.
Though our prototype, built using the Akka framework\footnote{Akka concurrency framework: \url{http://bit.ly/1B6WXGG}}, is designed to be distributed across multiple hosts, in the current implementation input graph partitionings reside on a single host.
Partitions are defined in terms of \textit{vertex-cut}, as opposed to edge-cut: \textit{inter}-partition connections are represented by flagging cut vertices and annotating them with the partitions they belong to.
We have extended Tinkerpop so that multiple edge-disjoint subgraphs are treated as a single, global, graph and queried using the Gremlin query language\footnote{The Gremlin query  language: \url{http://bit.ly/1tqUpWk}}.
An inter-partition traversal is detected when a Gremlin query retrieves the  external neighbours of a cut vertex.
Our test architecture is shown in Fig.~\ref{fig:architecture}.
It simulates a distributed deployment, where each partition is logically isolated,  managed by a separate instance of the \TAPR{} algorithm implementation. Each instance is responsible for updating the Visitor Matrix for its partition, and also determines the rank of extroverted vertices to evict at each iteration.
%
%
\subsection{Reducing the cost of the Visitor matrix}\label{subsection:space-complexity-heuristics}
As noted in Sec.~\ref{sec:visitor-matrix}, the space complexity of the VM for each partition $S_i$ grows with the number of vertices in the partition, and exponentially with the length of the query patterns: $O(|V_i|^t)$.
Here we discuss two heuristics, aimed at reducing the portion of the VMs that need to be explicitly represented or computed for each partition, reducing both the time and space complexity of the \TAPR{} algorithm.
\subsubsection{Space complexity}
Firstly, we note that large graphs are typically sparse 
: i.e. $|E| << |V|^2$.
As each vertex is only connected to a small number of neighbours, the adjacency and transition matrices representing such graphs contain many 0-value elements,
which may be discarded, compressing the matrices.
A VM, which is essentially a family of $k$ dimensional transition matrices where $2\leq k\leq t$ and $t$ is the number of traversal steps we remember, can be compressed using this standard technique.
Although in general we cannot be certain that the graphs against which \TAPR{} is applied will be sparse, the only non-zero elements that may exist in a VM are those that correspond to label paths in the pattern summary trie.
This serves to make the VM sparser relative to the corresponding adjacency matrix, \textbf{especially} well suited to compression.
Secondly, we avoid the costly computation and storage of many VM rows associated with vertices likely to be ``safe''; i.e. vertices unlikely to have high \textit{extroversion}.
Remember that, with \TAPR{}, we are only interested in identifying \textit{highly} extroverted vertices.
These are the most likely to be the source of \textit{inter}-partition traversals and therefore good candidates for being swapped to another partition.
From equations \ref{eq:introv} \& \ref{eq:extrov} (Sec.~\ref{subsection:introversion-extroversion}), we know that such extroverted vertices will necessarily have a low total \textit{intra}-partition traversal probability: low \textit{introversion}.
We therefore declare vertices with \textit{introversion} above a configurable threshold ``safe'' and discard them, reducing the space complexity of the VM.
Consider for example vertex 3 (denoted $v_3$) of partition $B$ in Fig.~\ref{fig:ex-graph-1}.
Accounting for the \textit{TPSTry} of Fig.~\ref{fig:ex-trie-1}, the traversal probabilities for $v_3$ are found in $\VMT_B(3)$ rows $\vmit{B}{2}[3,*]$, $\vmit{B}{3}[5,3,*]$, $\vmit{B}{3}[6,3,*]$ and so on.
The probability to be in $v_3$ from vertex 5 is computed as $\vmit{B}{1}[5]\cdot \vmit{B}{2}[5,3]$.
Extending this, the total intra-partition traversal probability \textbf{from} 3, given 5, is
\begin{displaymath}
	\vmit{B}{1}[5]\cdot \vmit{B}{2}[5,3] \cdot \underset{j}{\sum} \vmit{B}{3}[5,3,j]
\end{displaymath}

Given the values in Fig. \ref{fig:visitor-matrix}, completing this process for paths $p \in paths(v_3, V_B)$ to all $j \in V_B$ gives $v_3$ an \textit{intra}-partition traversal probability of 0.44.
Doing the same for all $j \in V$ gives a \textbf{total} traversal probability through $v_3$ ($Pr(v_3)$) of 0.5.
For any choice of \textit{introversion} threshold less than $\frac{0.44}{0.5} = 0.88$, $v_3$ would be a safe vertex.
We may discard any VM rows associated with $v_3$ except where necessary for paths through other, more extroverted, vertices.
%
%
\subsubsection{Time complexity}
In order to maximise the savings of the heuristic above, we would like to avoid computing some of the matrix rows we eventually discard.
We rely upon the following observations to achieve this:
as the probability of any given traversal from a vertex is usually less than $1.0$, longer paths of traversals generally have a lower probability than shorter ones;
the less likely a path of traversals though a vertex $v$, the less it will contribute to $v$'s \textit{introversion} and \textit{extroversion};
and the VM rows for each vertex $v$ are computed in ascending order of the length of their associated paths (Sec.~\ref{subsection:computing-vm-cells}).
Given these observations, we know that for the set of VM rows associated with a given vertex $v$:
those rows computed earlier \textbf{should} contribute more to $v$'s \textit{introversion} and \textit{extroversion} than those compute later.
We may therefore compare $v$'s \textit{introversion} to our chosen ``safe'' threshold after only having considered paths through $v$ of length up to $k$, where $k$ is less than the maximum length $k<t$.
We then do not need compute further VM rows for safe vertices.
In effect this provides another configurable threshold, this time controlling time complexity at the potential expense of accuracy.
The smaller the value of $k$ the more likely the algorithm is to declare a vertex safe which actually has a \textbf{total} \textit{introversion} below the ``safe'' threshold and might therefore have been an effective candidate for swapping to another partition.
Vertices without external neighbours represent a special case of this heuristic.
They are guaranteed to be ``safe'' and have no \textit{extroversion}. We do not calculate VM rows associated with these vertices, except where needed by other paths.
\subsection{TPSTry Implementation}\label{subsection:implementing-tpstries}
\TAPR{} captures the common query patterns in a workload stream, along with each pattern's probability, in the traversal pattern summary trie, \textit{TPSTry} (Sec.~\ref{sec:trie-encoding}).
\textit{TPSTry} is actually implemented as two separate data structures: i) a trie multimap, where each trie node maps to the set of queries which could be responsible for a traversal path with the associated sequence of vertex labels; and ii) a sorted table mapping queries to their respective frequencies.
These frequencies are approximated using a sketch datastructure
 which samples the occurrences of each query within a sliding window of time $t$.
\subsection{Calculating a partial extroversion order}\label{subsection:implementing-visitor-matrix}
\TAPR{} relies upon an ordering of the vertices in a partition by their likelihood to be the source of \textit{inter}-partition traversals.
In order to produces this order, we group the rows of a partition's visitor matrix by the final vertex of the paths they represent and then derive their \textit{extroversion} (Sec.~\ref{subsection:introversion-extroversion}).
As a result of the heuristics defined above, not all vertices are represented in the visitor matrix.
Therefore we refer to the sorted set of vertices produced as a \textit{partial extroversion ordering}.
%
%

%
Rather than grouping the rows of a pre-existing matrix, we define a corecursive algorithm to efficiently produce such rows consecutively during VM construction.
This greatly simplifies the process of maintaining a running total of \textit{intra}-partition transition probabilities for each vertex, as required for the heuristics presented in section \ref{subsection:space-complexity-heuristics}.
A simplified version of the procedure is expressed in Alg. \ref{algorithm:calculate-introversion}.
\begin{algorithm}
\small
\caption{Calculate the $\mathit{VM}_i$ rows for a vertex $v$}
\label{algorithm:calculate-introversion}
\begin{algorithmic}
\STATE $path \gets$ sequence of vertices (initially $(v)$)
\STATE $paths \gets$ set of paths (initially $\{(v)\}$)
\STATE $transitions \gets$ vector of probability values for $v$'s $neighbours$
\STATE $trie \gets$ traversal pattern summary trie
\STATE $threshold \gets$ safe introversion value
\STATE $length \gets$ max length of a path in $trie$
\STATE $rows\gets$ map of $path\rightarrow transitions$ vectors
\STATE $introversion(rows)\gets $ total \textit{introversion} of a set of $\mathit{VM}_i$ rows
\\ \hfill
\STATE $\mathbf{calcVMRows(paths, transitions, rows)}$
	\INDSTATE $newPaths\gets \emptyset$
	\INDSTATE $\mathbf{for}$ $path$ in $paths$ $\mathbf{do}$
		\INDSTATE[2] $\mathbf{if}$ $path$ size $>$ $length$ $\mathbf{then}$
$\mathbf{return}$ $rows$
		\INDSTATE[2] $\mathbf{if}$ $path$ in $trie$ $\mathbf{then}$
			\INDSTATE[3] $transitions\gets $ probabilities from $trie$ given $path$
			\INDSTATE[3] $rows\gets rows\ +$ ($path\rightarrow transitions$)
			\INDSTATE[3] $\mathbf{if}$ $introversion(rows)$ $> threshold$ $\mathbf{then}$
				\INDSTATE[4] $rows\gets \emptyset$
				\INDSTATE[4] $\mathbf{stop}$ $\mathbf{calcVMRows}$
		\INDSTATE[2] $neighbours\gets path.head.neighbours$
		\INDSTATE[2] $\mathbf{for}$ $n$ in $neighbours$ $\mathbf{do}$
			\INDSTATE[3] $newPaths\gets newPaths + n$ prepended to $path$
   	\RETURN $\mathbf{calcVMRows(newPaths, transitions, rows)}$
\end{algorithmic}
\end{algorithm}
Consider again our earlier example of vertex 3 in partition $B$ (Fig.~\ref{fig:ex-graph-1}), along with the pattern summary trie in Fig.~\ref{fig:visitor-matrix}(right).
Vertex 3 has the label $c$, which does exist as a prefix in the trie.
It has local neighbours 5 and 6, along with external neighbours 2 and 4, labelled $c$,$a$,$b$ and $d$ respectively.
The external transition probability from 3 given a path of $(3)$ is $1 - \Sigma (0, 1, 0)$ multiplied by the probability to have made the sequence of traversals which that path represents ($\frac{0.25}{|c|} = 0.125$). In this case: $0$.
The prefixes $cc$ and $ac$ also exist in the trie, therefore $(5,3)$ and $(6,3)$ are further potential paths through $3$.
The external transition probabilities from 3 given paths of $(5,3)$ and $(6,3)$ are $(1-\Sigma (0, 0, 1))\cdot 0.125$ and $(1-\Sigma (0, 0.25, 0.5))\cdot 0.25$ respectively.
Note that the total probability for the path $(6,3,j)$, $j\in V_B$ is $<1$ because $acd$ is also a prefix in the pattern summary trie, and vertex 3 is adjacent to the ``external'' vertex 4.
The final external transition probability from 3 is $0.06$; its \textit{extroversion} $0.06 \cdot \frac{1}{0.5} = 0.12$.
\subsection{Vertex Swapping}\label{subsection:vertex-swapping}
\begin{figure}
	\centering
	\includegraphics[scale=0.4]{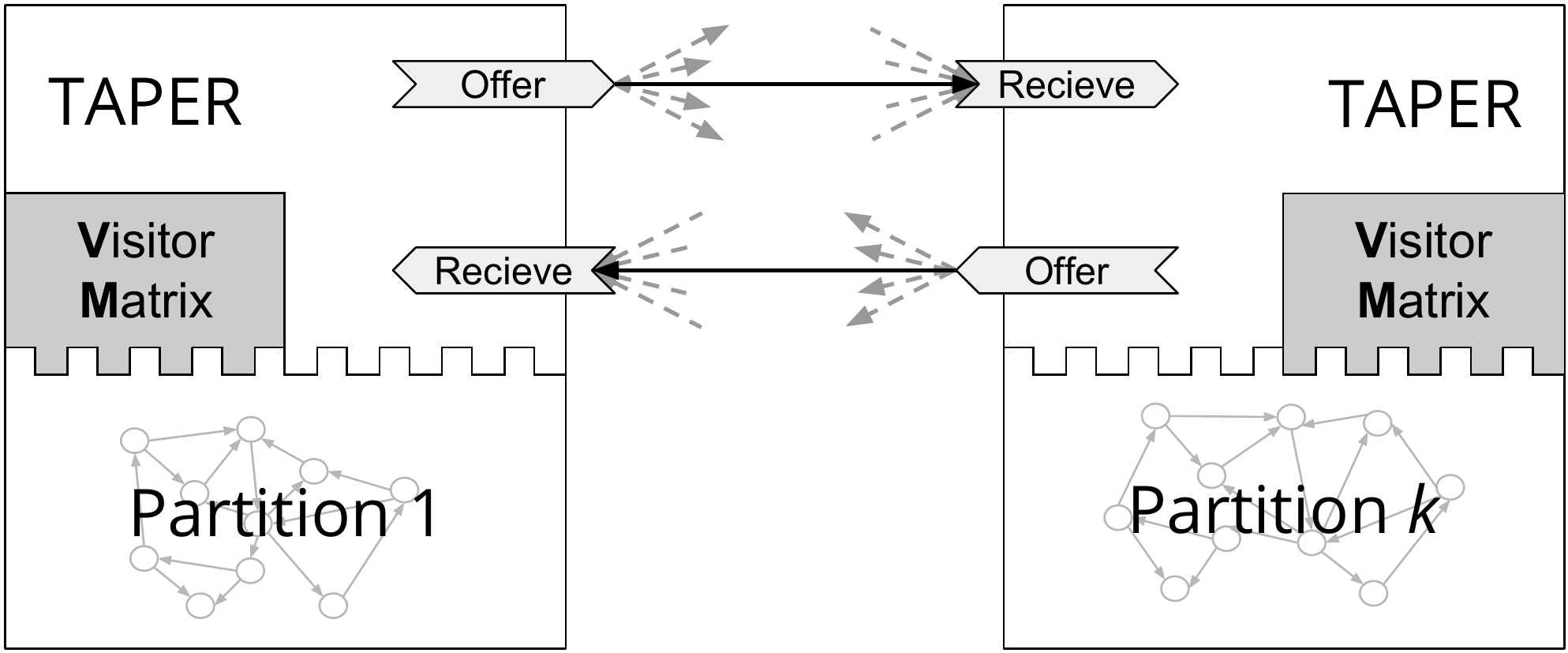}
	\caption{Offer/Receive algorithm in each \TAPR{} instance}
	\label{figure:offer-receive}
\end{figure}

To achieve its aims, \TAPR{} improves distributed query performance by reducing the probability of inter-partition traversals when answering queries.
For each partition, given a sorted collection of the vertices with the highest \textit{extroversion}, \TAPR{} must reduce this probability without mutating the underlying graph structure.
To achieve this we propose a simple variation on the k-way Kernighan-Lin algorithm proposed by Karypis and Kumar \cite{Karypis1998a} (Sec.~\ref{subsection:partition-enhancement}).
This is a two-step, symmetric process, shown in Fig.~\ref{figure:offer-receive}: firstly, given a priority queue of candidate vertices with high \textit{extroversion}, compute the preferred destination partition for each vertex, along with the clique of neighbours which should accompany it (its \textit{family}); secondly, when offered a new group of vertices, a partition should compute potential gains in introversion and decide whether or not to accept the offer 
.

We determine a swapping candidate's \textit{family} set with a simple recursive procedure:
Given each \textit{family} member (initially just the candidate), we examine its local neighbours; if a traversal from a neighbour to the member is more likely than not, then it is added to the \textit{family}. 
Once the \textit{family}-set has been determined, we evaluate the total loss in \textit{introversion} the sending partition would suffer from their loss.
%
%
This process is highly efficient as all the relevant values are preserved in the visitor matrix, either from calculating the introversion of vertices, or from constructing a candidates set in the previous step.
When on the receiving end of a swap, a partition should calculate the total local \textit{introversion} of a \textit{family}, and compare it to the potential loss to the offering partition.
Partitions are ``cooperative'' rather than greedy, so if the introversion gain for a receiving partition partition is not greater than the loss for a sending one, the swap is rejected.
%
%
If a swap is rejected the offering partition will try less ``prefereable'' destinations until all partitions adjacent to the candidate vertex are exhausted.
In this event the candidate vertex and its \textit{family} remain in their original partition.
This process runs independently for each partition, swapping \textit{extroverted} vertices to their preferred neighbouring partitions.
A vertex may only be swapped once per iteration of the algorithm.
When a partition's original queue of \textit{extroverted} vertices is empty, the resulting subGraph acts as input to a subsequent iteration of vertex-swapping.
Repeated iterations of this process will produce the desired result: an enhanced partitioning with a lower overall probability for inter-partition traversals, better \textit{workload-aware stability} given a query workload \Q{}.
%
%

\section{Evaluation}\label{section:evaluation}
Our evaluation aims to show how \TAPR{} achieves and maintains high partitioning quality, measured as low \textit{inter}-partition traversals, our proxy for high \textit{workload-aware stability} (Sec.~\ref{subsection:workload-aware-stability}).
We present three main results. Two on the effect of a single \TAPR{} invocation given a workload snapshot \Q{} (i.e. def.~\ref{eq:graph-repart}, from Sec.~\ref{subsection:taper-repart}):
%
\textbf{Firstly}, given a simple hash-partitioning  $P^0_k(G)$ and a workload \Q{}, a single \TAPR{} invocation (eqn.~\ref{eq:graph-repart}) achieves a quality comparable to that of a Metis-partitioning\cite{Karypis1998a} in at most 8 iterations;
\textbf{Secondly}, given the same workload, along with input partitionings generated by proven existing techniques, a \TAPR{} invocation is still able to achieve significant quality improvements.
Then one on the impact of a changing workload, given periodic \TAPR{} invocations (i.e. def.~\ref{eq:repeated-graph-repart} from Sec.~\ref{subsection:taper-repart}):
\textbf{Thirdly}, given a workload stream $\{ {\cal Q}_n\}$, our system maintains an upto-date query summary in the \textit{TPSTry}. As a result, repeated \TAPR{} invocations are able to keep $\ipt{}$ below some desired minimum, despite any workload changes.
We use Metis as our primary basis for comparison because, despite its age, it remains a gold standard for producing quality workload-agnostic partitionings of medium sized graphs \cite{Tsourakakis2012,Margo15}. As such it is a compelling yard-stick for our evaluation of the \TAPR{} prototype, which will consider partitioning quality only, not runtime performance. 
%

%
\subsection{Experimental setup}\label{subsection:test-methodology}

For all experiments we initially split the test graphs into a reasonable number of partitions (8), using either hash or Metis partitioning, as described below. 
As mentioned, we consider the quality of a partitioning as its workload-aware stability (Sec.~\ref{subsection:workload-aware-stability}), corresponding to the probability of inter-partition traversals when executing a workload ${\cal Q}$.
We measure this experimentally by executing snapshots of query workloads over partitioned graphs and counting (Sec.~\ref{section:implementation}) the number of inter-partition traversals $\ipt$.
All algorithms, data structures and dataset pre-processing steps, including \textit{calcVMRows} and the \textit{TPSTry}, 
are publicly available\footnote{The \TAPR{} repository: \url{http://bit.ly/1W3f0eH}}. 
All our experiments are performed on a machine with a 3.1Ghz Intel Core i7 CPU and 16GB of RAM.
\begin{figure*}
    \centering
    \begin{minipage}{0.70\textwidth}{

        \includegraphics[width=0.97\linewidth]{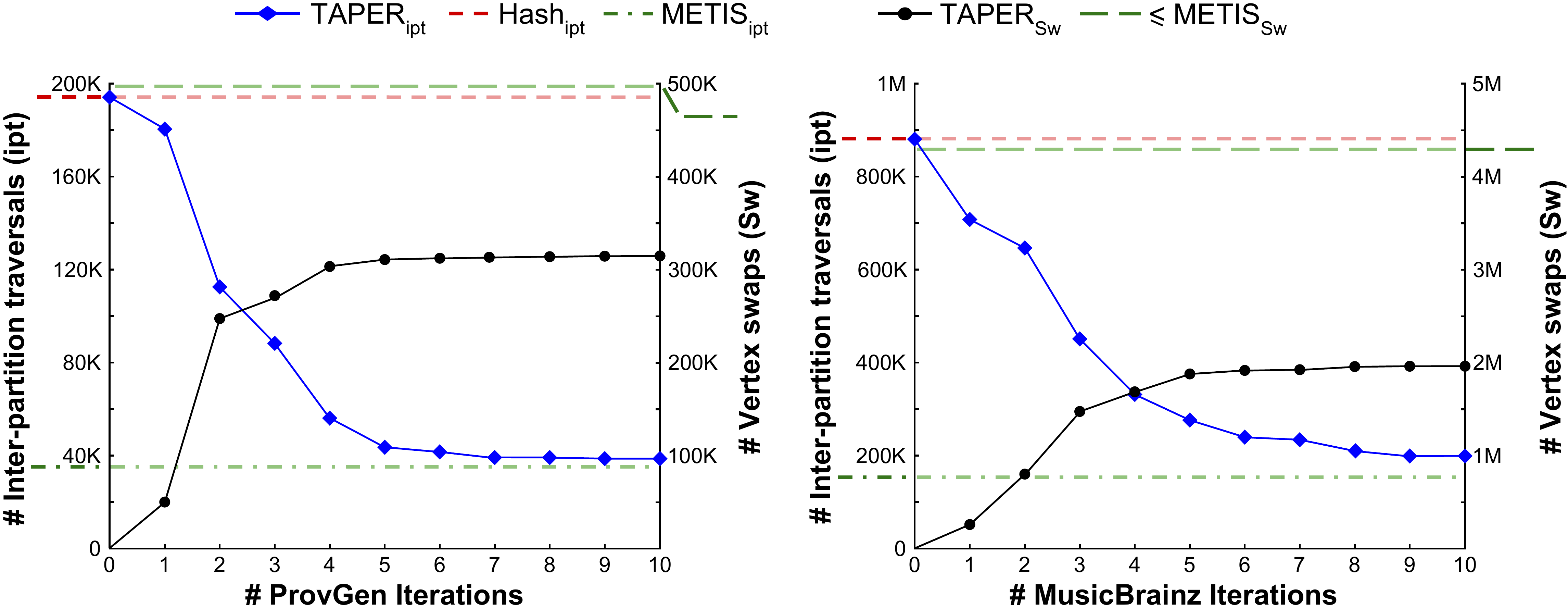}}
        \captionof{figure}{$\ipt$ per \TAPR{} internal iteration}
        \label{fig:traversals-vs-iterations}
    \end{minipage}%
    \begin{minipage}{0.30\textwidth}{
        \includegraphics[width=\linewidth]{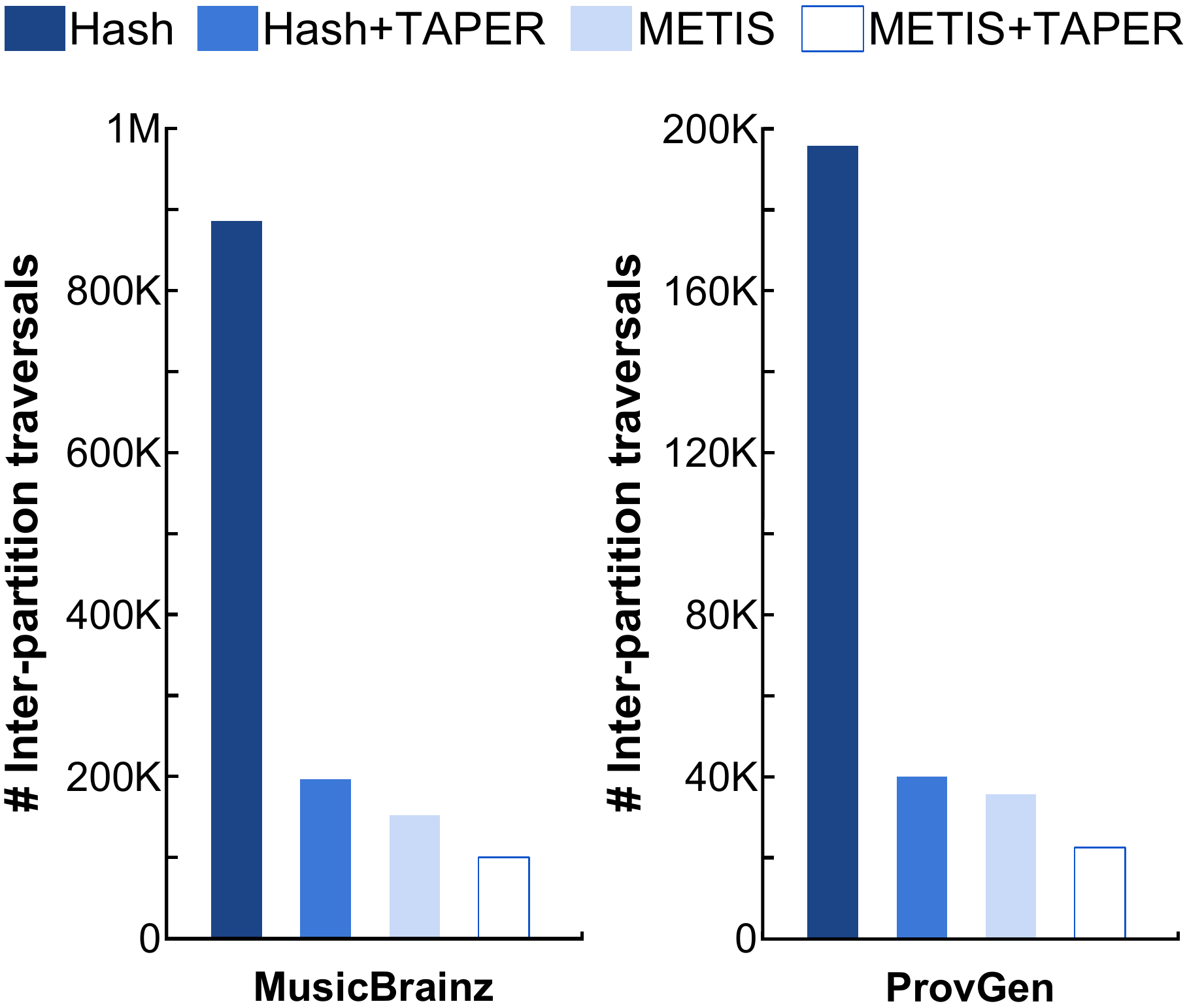}}
         \captionof{figure}{$\ipt$ per approach}
         \label{fig:traversals-vs-approach}
     \end{minipage}
\end{figure*}
\subsubsection{Test Datasets}\label{subsection:datasets}
\TAPR{} is designed to perform best on highly heterogeneous graphs, as noted in Sec. \ref{sec:trie-encoding}.
%
When the graph is homogeneous, the uniformity of traversal probabilities renders min edge-cut an equally good measure of partition quality.
Thus, we have tested the algorithm on two heterogeneous graphs.
The first, \textbf{MusicBrainz}\footnote{The MusicBrainz database: http://bit.ly/1J0wlNR}, is a freely available database of music which contains curated records of artists, their affiliations and their works.
This database currently stores over 950,000 artists and over 18 million tracks.
When converted to a graph, the subset of data we use amounts to around 10 million vertices and more than 30 million edges.
The graph is also highly heterogeneous, containing more than 12 distinct vertex labels.
The second test dataset is a synthetic provenance graph, generated using the \textbf{ProvGen} generator \cite{Firth2014} and compliant with the PROV data model~\cite{w3c-prov-dm}.
ProvGen is designed to produce arbitrarily large PROV graphs starting from small seed graphs and following a set of user-defined topological constraint rules.
Provenance graphs are a form of metadata, which contains records of the history of entities, e.g. documents, complex artifacts, etc\ldots
They are exemplars of the large-scale heterogeneous graphs that \TAPR{} is designed to partition.
For the purpose of these experiments we generated a graph with about 1 million vertices and 3 million edges.
As described in \cite{w3c-prov-dm}, PROV graphs naturally have three labels, representing the three main elements of provenance: $Entity$ (data), $Activity$ (the execution of a process that acts upon data), , and $Agent$, namely the people, or systems, that are responsible for data and activities.

\subsubsection{Test query workloads}\label{subsection:queries}
For each dataset we need to create a corresponding query stream: an infinite sequence of pattern matching queries consisting of a small number of distinct graph patterns.
The relative frequencies of each query pattern should shift continuously, representing workload changes with time.
For our experiments, we selected a simple periodic model of workload change where the frequency of each query pattern grows and shrinks according to a constant, repeating pattern\footnote{Similar to a sin wave. Further details of the workload stream are elided for space.} and no new query patterns are added over time. 
These frequency changes are the compliment of each other, so that the total frequency of all query patterns in the workload stream is always equal to 1.  
Note that \TAPR{} \textbf{does not assume} any such distribution of query frequencies, and can refine a graph partitioning given arbitrary changes in workload.
We also define the set of distinct query patterns for each dataset.
Regarding MusicBrainz, to the best of our knowledge there is no widely accepted corpus of benchmark queries.
Thus, we define a small set of common-sense queries that focus on discovering implicit relationships in the graph, such as collaborations between artists, and migrations between geographical areas.\medskip
$\mathbf{MQ_1}$ $Area\cdot Artist\cdot (Artist|Label)\cdot Area$: searches for two distinct patterns which would indicate an artist has moved away from their country of origin.
$\mathbf{MQ_2}$ $Artist\cdot Credit\cdot (Track|Recording)\cdot Credit\cdot Artist$: might be used to detect collaboration between 2 or more artists on a single track.
$\mathbf{MQ_3}$ $Artist\cdot Credit\cdot Track\cdot Medium$: would return a set of all the Mediums (e.g. Cd) which carry an Artist's work.\medskip

Regarding provenance graphs, several categories of typical pattern matching queries have been proposed \cite{Karvounarakis}.
Using these categories, we propose four query patterns typical of provenance analysis.\medskip
$\mathbf{PQ_1}$ $Entity\cdot (Entity)*\cdot Entity$: computes the transitive closure over a data derivation relationship.
$\mathbf{PQ_2}$ $Agent\cdot Activity\cdot Entity\cdot Entity\cdot Activity\cdot Agent$: identifies pairs of agents who  have collaborated as data producer/consumers pairs.
$\mathbf{PQ_3}$ $(Entity)*\cdot Activity\cdot Entity$: returns all entities and all activities involved in the creation of a given entity.
$\mathbf{PQ_4}$ $Entity\cdot Activity\cdot (Agent)*$: returns agents responsible for the creation of a given an entity.
\subsection{Results}

\begin{figure*}
    \centering
    \begin{minipage}{0.26\textwidth}{
        \includegraphics[width=\linewidth]{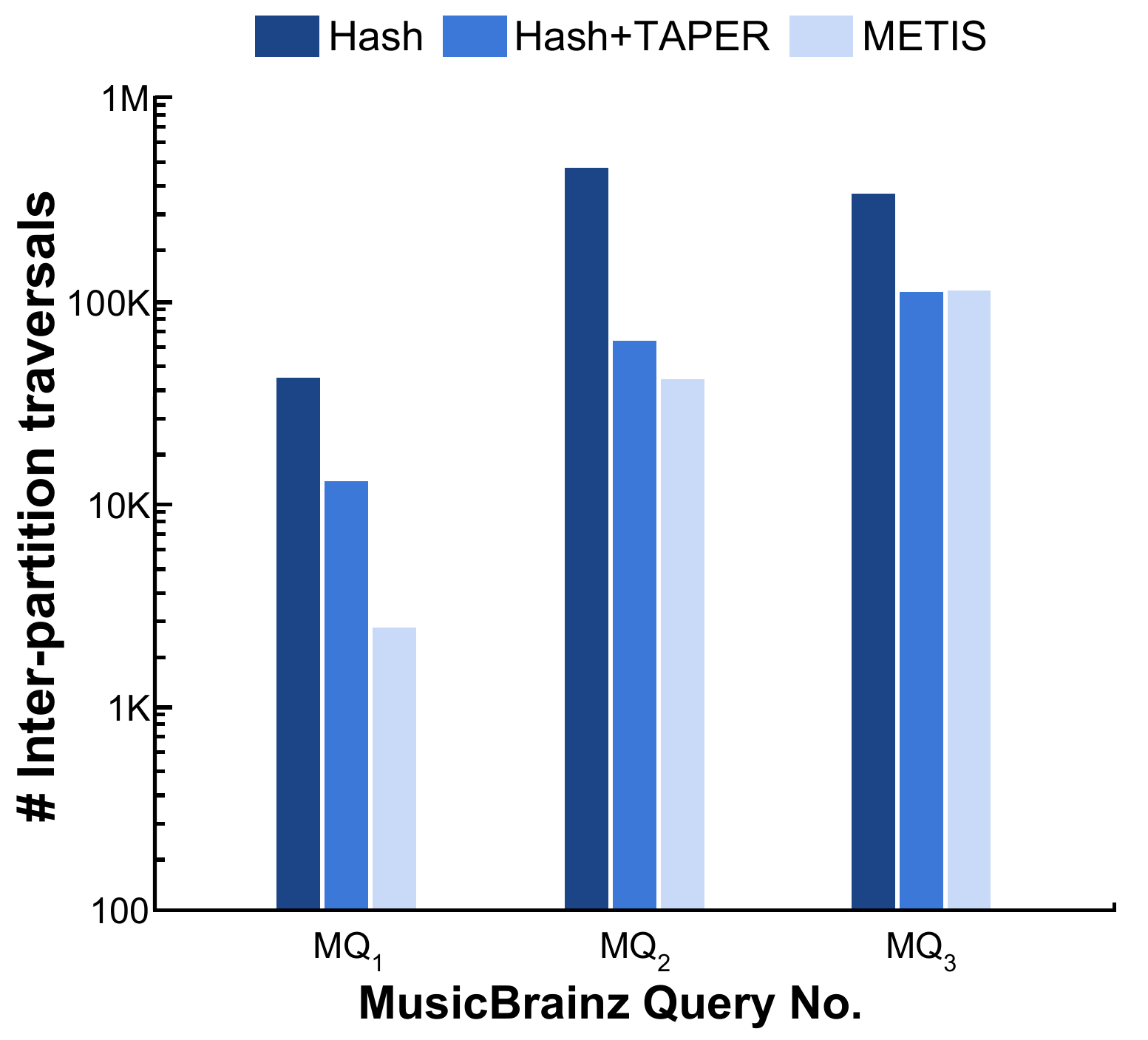}}
        \captionof{figure}{$\ipt$ per query}
        \label{fig:xtraversal-reduction-pct}
    \end{minipage}%
    \quad
    \begin{minipage}{0.29\textwidth}{
        \includegraphics[width=0.96\linewidth]{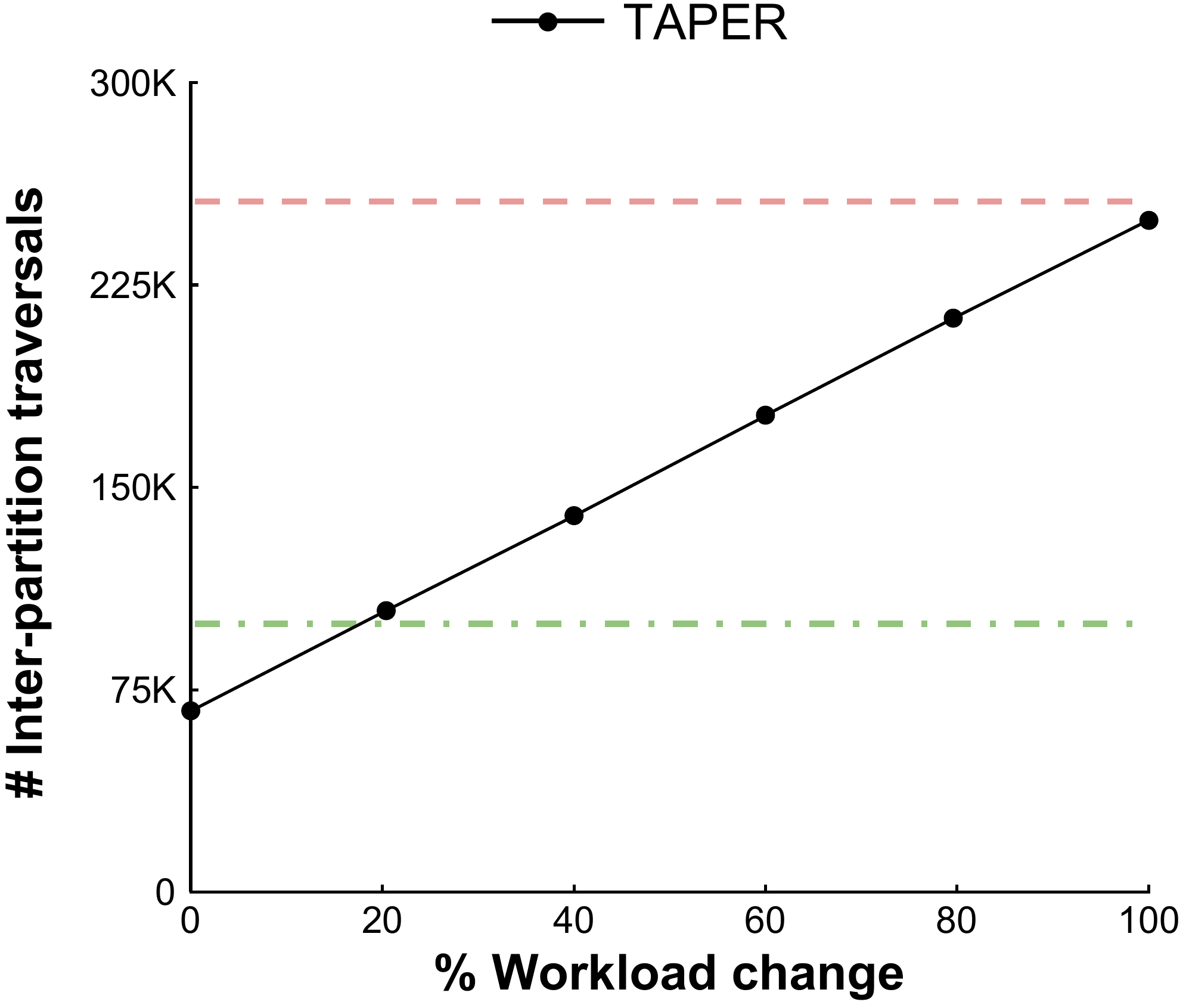}}
        \captionof{figure}{$\ipt$ vs Workload change}
        \label{fig:xtraversal-vs-workload-change}
    \end{minipage}%
    \quad
    \begin{minipage}{0.40\textwidth}{
        \includegraphics[width=\linewidth]{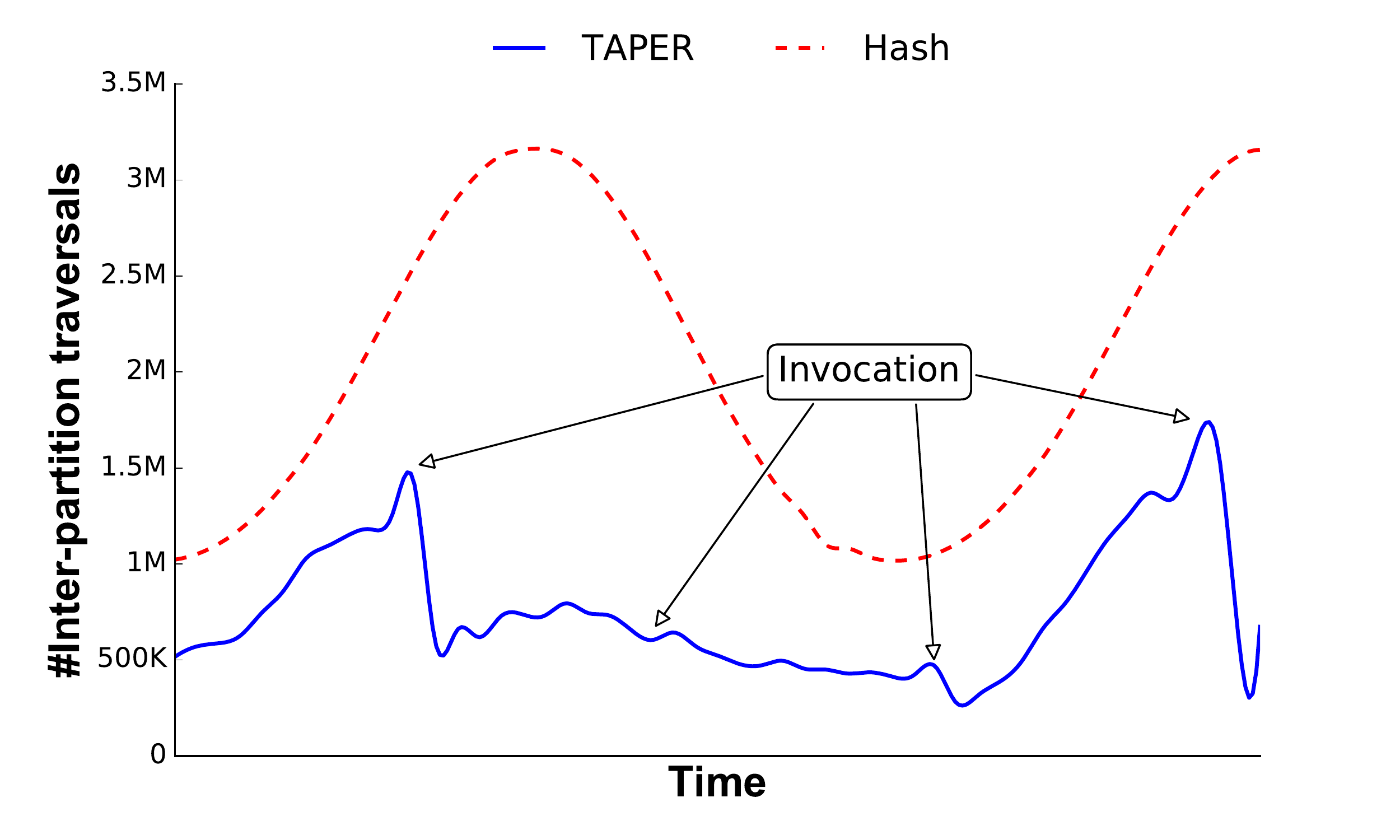}}
        \captionof{figure}{$\ipt$ over time w. \TAPR{} invocations}
        \label{fig:xtraversal-w-repeated-taper}
    \end{minipage}%
\end{figure*}

\subsubsection{Improvement over an initial hash partitioning}\label{subsection:effects-of-enhancement}
Fig.~\ref{fig:traversals-vs-iterations} shows the improvement in partitioning quality which a single \TAPR{} invocation achieves for each dataset, given static workload snapshots ${\cal Q}$ and initial hash-partitionings $P^0_8(G)$.
The top dotted line bisecting the left y-axis is our baseline: the $\ipt$ required to execute ${\cal Q}$ over $P^0_8(G)$.
The bottom dotted line indicates the $\ipt$ required to execute ${\cal Q}$ over an initial Metis partitioning.
The chart shows how partitioning quality converges to within $10\%$ of that over a Metis partitioned graph, after fewer than 8 internal iterations.
Note that these iterations also satisfy a maximum partition imbalance of $5\%$.
Figure \ref{fig:traversals-vs-iterations} also demonstrates that a \TAPR{} invocation requires far less communication than a full Metis repartitioning.
5 iterations of \TAPR{} over the ProvGen dataset (Fig.~\ref{fig:traversals-vs-iterations}(a)) required 300k vertex swaps to produce its $\sim 80\%$, enhancement.
On the other hand, the number of swaps required to rearrange Hash partitions to be consistent with a Metis partitioning (i.e the cost of a Metis repartitioning) is at least 500k.
This cost also ignores any communication cost associated with actually computing the repartitioning, such as gathering the graph ($|V|$ swaps).
Practically, a Metis repartitioning has a cost \textbf{at least 2X} that of a \TAPR{} invocation in both our test cases, yet achieves only a small improvement in query performance.
This suggests that, by performing swaps in \textit{extroversion} order (Sec.~\ref{subsection:partition-enhancement}), we are correctly prioritising those swaps that are more effective at reducing $\ipt$, given a workload snapshot ${\cal Q}$.
This supports \TAPR{}'s applicability to continuous re-partitioning in online settings, such as distributed graph DBMS, where other system requirements may severely limit the number of vertex swaps possible in a given timeframe.
\subsubsection{Improving over other initial partitionings}\label{subsection:effects-of-enhancement-plus-metis}
Fig \ref{fig:traversals-vs-approach} illustrates that a \TAPR{} invocation may achieve a quality improvement over not only an initial hash-partitioning, but also over initial partitionings produced with existing partitioning techniques, including 
Metis.
When improving upon a Metis partitioning (METIS + \TAPR{} in the figure), \TAPR{} achieves an average $30\%$ reduction in $\ipt$
.
As seen in the previous section, a \TAPR{} invocation over an initial hash-partitioning achieves a quality less than an initial Metis partitioning.
Thus we conjecture that the \TAPR{} algorithm is sensitive to its starting input and, despite swapping vertex \textit{family} cliques (Sec.~\ref{subsection:vertex-swapping}), when starting from a Hash partitioning is gets trapped in local optimisation minima.
Starting from a Metis partitioning, \TAPR{} iteratively approaches a new minimum closer to the global.
\TAPR{}'s ability to improve over Metis graphs my be explained by observing that in non-trivial partitionings, some edges must cross partition boundaries.
As a workload-agnostic algorithm, Metis is optimising for a different cost function than \TAPR{} and may cut edges which are likely to be frequently traversed, giving \TAPR{} scope for its improvement.
Note that improvement is not necessarily possible when Metis is given an input graph with edge-weights corresponding to traversal likelihood given Q.
In that instance, edge-weight cut is equivalent to inter-partition traversal probability: both Metis and \TAPR{} are optimising for the same cost function.
However, tracking workload with edge-weights is challenging, and adapting to workload changes with Metis would still require a full repartitioning.
\subsubsection{Optimising for frequent queries}\label{subsection:optimising-for-frequent-queries}
%

%
%

%
Fig. \ref{fig:xtraversal-reduction-pct} demonstrates the effect of \TAPR{}'s use of query frequencies within a workload to prioritise vertex swaps.
The figure presents $\ipt$ over various partitionings of the Musicbrainz graph, given a workload snapshot with the relative frequencies of queries $\mathbf{MQ_1}$, $\mathbf{MQ_2}$, and $\mathbf{MQ_3}$ at $10\%$, $20\%$ and $70\%$, respectively.
Relative to the Metis partitioning, a \TAPR{} invocation achieves its worst quality for $\mathbf{MQ_1}$, improving with $\mathbf{MQ_2}$ and surpassing the other system for $\mathbf{MQ_3}$.
This is because paths in a graph which form a full, or partial, match of a high frequency query afford their vertices and edges a higher probability of being traversed.
When edges in the path cross partition boundaries, this traversal probability contributes to \textit{extroversion}.
Again, \TAPR{} is prioritising vertex-swaps to internalise paths traversed by the most common queries to single partitions.
\subsubsection{The effect of changes in query workloads}\label{subsection:workload-changes}
So far in our evaluation we have performed single invocations of TAPER: several iterations of vertex-swapping over an initial partitioning, given a static snapshot of queries. 
This is essentially \textbf{fitting} the distribution of vertices across partitions to a particular workload snapshot's dominant traversal patterns (Sec.~\ref{subsection:workload-aware-stability}).
However, within a larger workload stream, query frequencies are likely to change continuously. 
Fig \ref{fig:xtraversal-vs-workload-change} trivially demonstrates that the quality of a fitted partitioning degrades in the presence of such workload change.
%

%
%

%
For clarity this experiment was performed over the provenance dataset, with a finite workload stream comprised of two query patterns: $\mathbf{Q_a}\ Entity\cdot Entity$  and $\mathbf{Q_b}\ Agent\cdot Activity$. 
At the head of the stream, the frequency of $\mathbf{Q_a}$ is $100\%$; throughout the stream the frequency of $\mathbf{Q_a}$ tends linearly to $0\%$, $\mathbf{Q_b}$ to $100\%$.
The initial partitioning has been pre-improved with \TAPR{}, assuming a workload of $100\%$ $\mathbf{Q_a}$ queries.
As the frequency of $\mathbf{Q_b}$ queries increases, so does the $\ipt$. 
For comparison, the top dotted line in Fig \ref{fig:xtraversal-vs-workload-change} shows the $\ipt$ required to execute solely $\mathbf{Q_b}$ queries over a hash-partitioning of the graph; the bottom line shows those required over a partitioning improved by \TAPR{} \textbf{correctly assuming} $\mathbf{Q_b} = 100\%$.
In other words, in the presence of an unexpected change in workload, \TAPR{}'s quality improvement may degrade to near that of a naive hash-partitioner.
However, the \textit{TPSTry} is continually updated to reflect changing query frequencies (Sec.\ref{sec:trie-encoding}) and our experiments depicted in fig \ref{fig:traversals-vs-iterations} demonstrate that \TAPR{} invocations are inexpensive compared to a full re-partitioning operation.
Therefore, by periodically executing \TAPR{} invocations with the current partitioning as input, we are able to maintain our partitioning quality improvement even in the presence of a dynamic and changing workload stream.
Fig \ref{fig:xtraversal-w-repeated-taper} presents the $\ipt$ which occur when executing a full streaming query workload, generated as described (Sec.~\ref{subsection:queries}), over the Musicbrainz graph partitioning. 
%
%
Knowing the $\ipt$ required to execute each query pattern over a hash partitioning, the chart displays a derived trendline for baseline performance.
As the frequency of queries which return more results rises, then falls, the $\ipt$ follows suit.   
Comparing against this baseline, Fig.~\ref{fig:xtraversal-w-repeated-taper} clearly demonstrates that, with periodic invocations, \TAPR{} is able to prevent some performance decay over time.
The highlighted areas of the chart indicate when \TAPR{} has been executed; each followed by a drop in $\ipt$, as we expect.
In this experiment we trigger \TAPR{}'s execution at regular intervals, which is naive, as invocations may occur when a trend in the workload renders them unnecessary or detrimental.
For instance, the second highlighted invocation acts on stale information and actually causes a slight \textit{rise} in $\ipt$.
Identifying effective trigger conditions is left as future work.
%
%

\section{Conclusions}
In this paper, we have presented \TAPR{}: a practical system for improving path query processing performance in partitioned graph data.
By monitoring the traversals and frequencies associated with queries in a workload stream, we can calculate the likelihood for any vertex in a graph to be a source of costly inter-partition traversals - its \textit{extroversion}.
By using vertex labels as an intensional representation of traversal patterns, along with several other heuristics, the resource intensive challenge of of identifying and relocating these most \textit{extroverted} vertices becomes tractable.
Our experiments show that \TAPR{} significantly reduces the number of inter-partition traversals ($\ipt$) over a graph partitioning.
It achieves improvements similar to quality existing partitioners, such as Metis
, whilst requiring a lower total communication volume, even after many internal iterations of its vertex-swapping algorithm.
Furthermore, as it is workload-aware, \TAPR{} may even improve the quality of input partitionings already good \textit{w.r.t} some  workload agnostic objective function, such as min edge-cut.
We discuss how we maintain the \textit{TPSTry} datastructure given a stream of pattern matching queries (Sec~\ref{sec:trie-encoding}), summarising probable traversal patterns at any time.
Our final experiments show that as a result of this continuous summary, and the incremental nature of \TAPR{} invocations, we are able to maintain high partitioning quality (\textit{workload-aware stability}) in the presence of a changing stream of queries.
This renders \TAPR{} suitable for use in online scenarios, where such dynamic workloads are common.
As mentioned (Sec.~\ref{section:definitions}), in \TAPR{} we only consider queries as single regular expressions over vertex labels when encoding traversal patterns in the \textit{TPSTry}.
For future work we plan to extend our workload summary, including edge labels and more complex query patterns.
This will increase the accuracy of extroversion orderings, improving performance.
We also plan to explore more sophisticated, predictive, trigger conditions for \TAPR{} invocations when given a workload stream, as the current regular intervals are ineffective.
%
%


\bibliographystyle{abbrv}
\enlargethispage{\baselineskip}
\small
\bibliography{tapr}

\end{document}